\documentclass[final,3p,times,twocolumn]{elsarticle}
 \biboptions{comma,sort&compress}

\usepackage{graphicx}
\usepackage{here}
\usepackage{cuted}
\usepackage{amssymb}
\usepackage[fleqn]{amsmath}
\usepackage[figuresright]{rotating}

\newcommand{\lapprox}{%
\mathrel{%
\setbox0=\hbox{$<$}
\raise0.6ex\copy0\kern-\wd0
\lower0.65ex\hbox{$\sim$}
}}
\newcommand{\gapprox}{%
\mathrel{%
\setbox0=\hbox{$>$}
\raise0.6ex\copy0\kern-\wd0
\lower0.65ex\hbox{$\sim$}
}}
\newcommand{\be}{\begin{equation}}
\newcommand{\ee}{\end{equation}}
\newcommand{\bea}{\begin{eqnarray}}
\newcommand{\eea}{\end{eqnarray}}

\newcommand{\lbl}[1]{\label{eq:#1}}
\newcommand{ \rf}[1]{(\ref{eq:#1})}

\def\beq{\begin{equation}}
\def\eeq{\end{equation}}
\def\bea{\begin{eqnarray}}
\def\eea{\end{eqnarray}}
\def\bq{\begin{quote}}
\def\eq{\end{quote}}

\def\nnb{\nonumber}
\def\ga{\left(}
\def\dr{\right)}

\def\lrar{\Longrightarrow}
		
\def\nnb{\nonumber}
\def\la{\langle}
\def\ra{\rangle}
\def\nin{\noindent}
\def\ba{\vspace*{-0.2cm}\begin{array}}
\def\ea{\end{array}\vspace*{-0.2cm}}

\def\b{$\bullet~$}
\def\als{\alpha_s}

\def\gg2{ \la\alpha_s G^2 \ra}
\def\gg3{g^3f_{abc}\la G^aG^bG^c \ra}
\def\ggg4{\la\als^2G^4\ra}

\begin{document}

\begin{frontmatter}

%%
%%%%%%%%%%%%%%%%%%%%%%%%%%%%%%%%%%%%%%%%%%%%%%%%%
\title{ Scalar Meson Contributions to $a_\mu$ from  Hadronic Light-by-Light Scattering  }
% \corref{cor1}
% }
% \cortext[cor1]
 \author[label1]{M. Knecht}
%  \cortext[cor0]{FAPESP CNPq-Brasil PhD student fellow.}
\ead{marc.knecht@cpt.univ-mrs.fr}
\address[label1]{Centre de Physique Th\'{e}orique UMR 7332, CNRS/Aix-Marseille Univ./Univ. du Sud Toulon-Var \\
CNRS Luminy Case 907, 13288 Marseille Cedex 9, France}
% \author[label2]{F. Fanomezana  \fnref{fn0} }
% \corref{cor0}}
 % \cortext[cor0]
%  \fntext[fn0] {PhD student.}

%\ead{fanfenos@yahoo.fr}
\author[label2]{S. Narison
\fnref{fn0}}
  \fntext[fn0]{Madagascar consultant of the Abdus Salam International Centre for Theoretical Physics (ICTP), via Beirut 6,34014 Trieste, Italy.}
    \ead{snarison@yahoo.fr}
    \address[label2]{Laboratoire
Particules et Univers de Montpellier, CNRS-IN2P3, 
Case 070, Place Eug\`ene
Bataillon, 34095 - Montpellier, France.}
 %   \address[label3a]{Speaker}
%\address[label4]{Madagascar consultant of the Abdus Salam International Centre for Theoretical Physics (ICTP), via Beirut 6,34014 Trieste, Italy .}
 \author[label3]{A. Rabemananjara}
 \address[label3]{Institute of High-Energy Physics (iHEPMAD), University of Antananarivo, 
Madagascar}
%  \cortext[cor2]{Ph.D. student}
\ead{achrisrab@gmail.com}

 \author[label3]{D. Rabetiarivony\fnref{fn1}}
  \fntext[fn1]{PhD student.}
\ead{rd.bidds@gmail.com} 

% \author[label2]{G.~Randriamanatrika\fnref{fn0}}
%  \cortext[cor2]{Ph.D. student}
%\ead{artesgaetan@gmail.com}

\pagestyle{myheadings}
\markright{ }
\begin{abstract}
Using an effective $\sigma/f_0(500)$ resonance, which describes the $\pi\pi\to\pi\pi$ and $\gamma\gamma\to\pi\pi$ scattering data,
we evaluate its contribution and the ones of the other  scalar mesons to the 
hadronic light-by-light (HLbL) scattering  component of the anomalous magnetic moment $a_\mu$ of the muon. We obtain the conservative range of values: 
$ \sum_Sa_\mu^{lbl}\vert_S\simeq -\ga 4.51\pm 4.12\dr \times 10^{-11}$, which is dominated by the $\sigma/f_0(500)$
contribution ($ 50\%\sim 98\%$), and where the large error is due to the uncertainties on the parametrisation of the  form factors. Considering our new result, 
we update the sum of the different theoretical contributions to
$a_\mu$ within the standard model, which we then compare to experiment. This comparison gives $(a_\mu^{\rm exp} - a_\mu^{\rm SM})= +(312.1\pm 64.6)\times 10^{-11}$, 
where the theoretical errors from HLbL are dominated by the scalar meson contributions. 
\end{abstract}
%%%%%%%%%%%%%%%%%%%%%%%%%%%%%%%%%%%%%%%%%%%%%%%%%%%%%%%%%
% \begin{document}

\end{frontmatter}
%{\Large {\bf I. Numerical Checks of the pion contribution to LBL}} 
%%%%%%%%%%%%%%%%%%%%%%%%%%%%%%%%%%%%%%
\section{Introduction}
The anomalous magnetic moments $a_\ell$ ($\ell \equiv  e , \mu$) of the light charged leptons, electron and muon, are among
the most accurately measured observables in particle physics. The relative precision achieved by the latest experiments 
to date is of 0.28 ppb in the case of the electron \cite{Odom06,Hanneke08}, and 0.54 ppm in the case of the muon
\cite{E82106}. An ongoing experiment at
Fermilab \cite{FNAL,Roberts09,Gorringe:2017bhq}, and a planned experiment at J-PARC \cite{JPARC}, 
aim at reducing the experimental uncertainty on $a_\mu$ to the level of 0.14 ppm,
and there is also room for future improvements on the precision of $a_e$.
The confrontation of these very accurate measurements with equally precise predictions
from the standard model then provides a stringent test of the latter, and, as the
experimental precision is further increasing, opens up the possibility of
indirectly revealing physics degrees of freedom that even go beyond it.

From this last point of view, the present situation remains unconclusive in the case of the muon
(in the case of the electron, the measured value of $a_e$ agreed with the predicted value 
obtained from the measurement of the fine-structure constant of Ref. \cite{Bouchendira:2010es};
however, the more recent determination of $\alpha$ \cite{Parker2018} now results in a tension
at the level of $2.5$ standard deviations between theory and experiment).
Indeed, the latest standard model evaluations of $a_\mu$
(Ref. \cite{Knecht:2018nhq} provides a recent overview, as well as references
to the literature; see also Section \ref{sec:a^lbl_values} at the end of this article) 
reveal a discrepancy between theory
and experiment, which however is at the level of $\sim 3.5$ standard deviations only.
It is therefores mandatory, as the experimental precision increases, to also
reduce the theoretical uncertainties in the evaluation of $a_\mu$.

%%%%%%%%%%%%%%%%%%%%%%%%%
\begin{figure}[b]
\begin{center}
{\includegraphics[width=2cm  ]{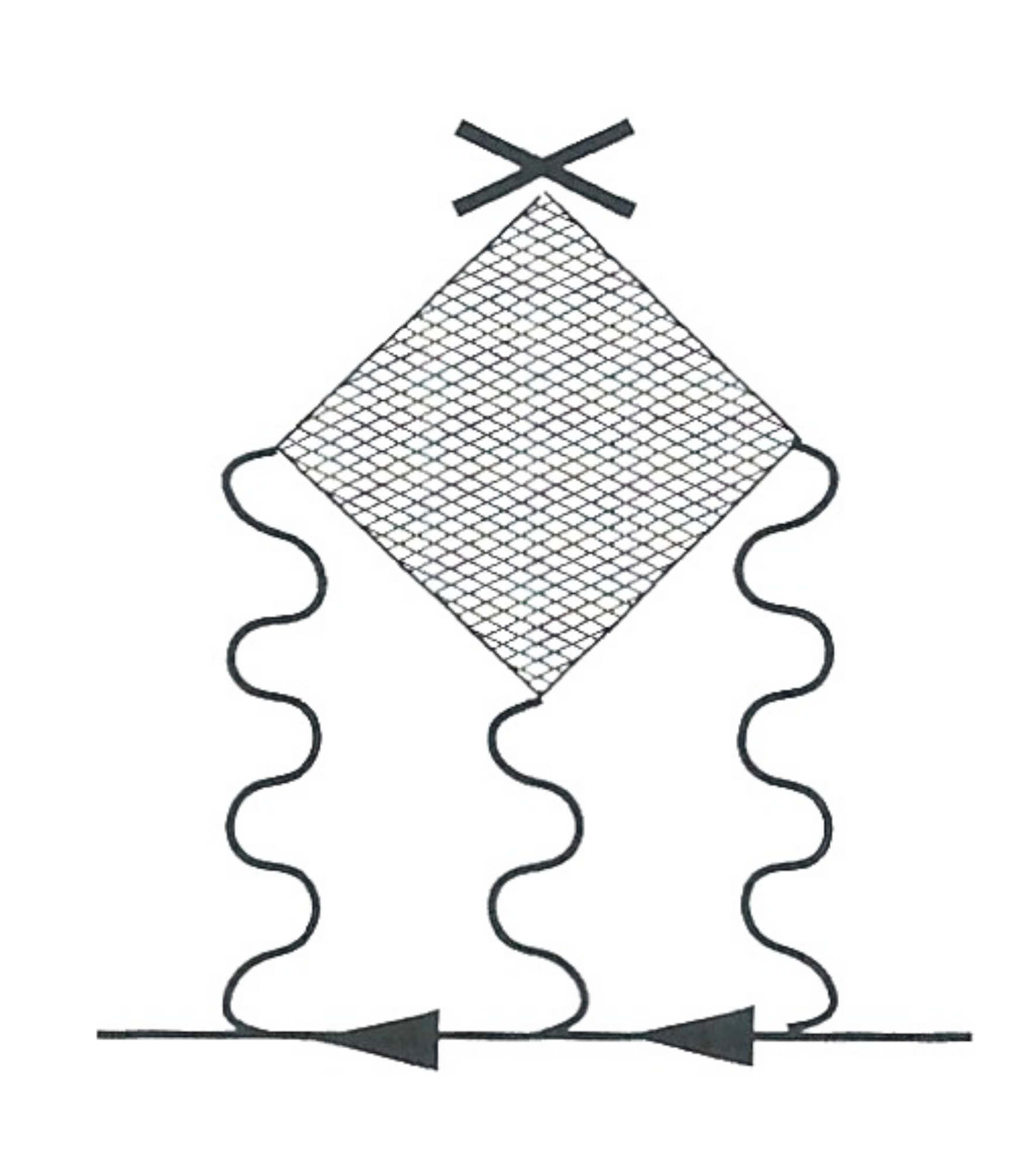}}\\
\caption{ Light-by-light Hadron scattering contribution to $a_l$. The wavy lines represent photon. The cross correponds to the insertion of the electromagnetic current. 
The shaded box represents hadrons subgraphs.}
\label{fig:box}
\end{center}
\end{figure}
 %%%%%%%%%%%%%%%%%%%%%%%%%%%%%%%%%%%%%%%

\begin{figure}[t]
\begin{center}
{\includegraphics[width=7cm  ]{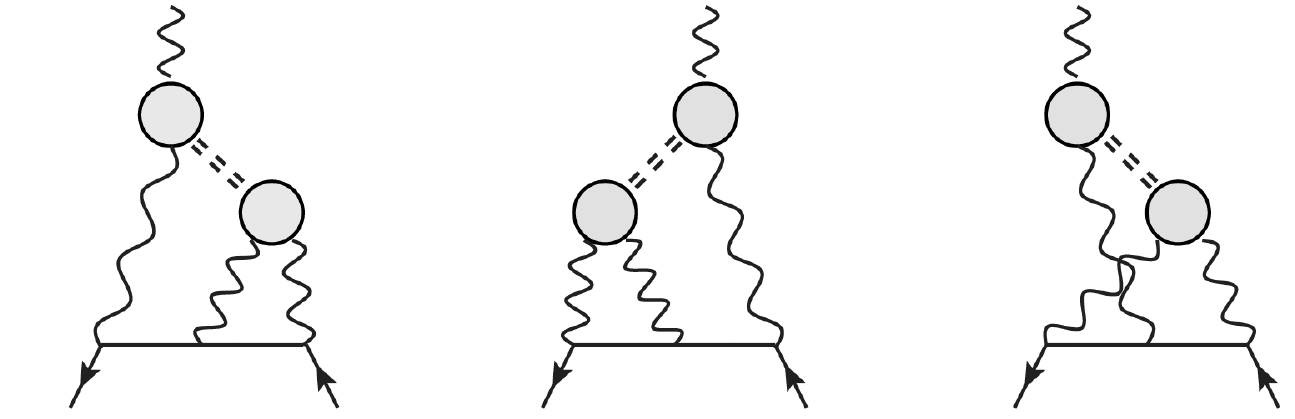}}\\
\hspace*{0.4cm} (a)\hspace*{2.1cm} (b)\hspace*{2.1cm} (c)
\caption{Scalar meson exchange (dotted lines) to  Light-by-light scattering contribution to $a_\mu$. The wavy lines represent photon. 
The shaded blob represents form factors. The first and second diagrams contribute to the function $T_1$, and the third to the function $T_2$ 
defined in Eq.\,\ref{eq:amu-mink}. }
\label{fig:scalar}
\end{center}
\end{figure}

Presently, the limitation in the theoretical precision of $a_\mu$ is due to the contributions
from the strong interactions, which are dominated by the low-energy, non perturbative, regime
of quantum chromodynamics (QCD). The present work is devoted to a hadronic contribution
arising at order ${\cal O} (\alpha^3)$, and currently refered to as hadronic light-by-light
(HLbL), see Fig. \ref{fig:box}. More precisely, we will be concerned with a particular
contribution to HLbL, due to the exchange of the $0^{++}$ scalar states
$\sigma/f_0 (500)$, $a_0 (980)$, $f_0 (980)$, $f_0 (1370)$, and $ f_0 (1500)$. In earlier evaluations
of the HLbL part of $a_\mu$, some of these states were either treated in the framework of the extended
Nambu--Jona-Lasinio model \cite{Bijnens:1995cc,Bijnens:1995xf}, or they were simply omitted altogether
\cite{Hayakawa:1995ps,Hayakawa:1996ki}. More recently, in Ref. \cite{BARTOS} the contributions from the 
$\sigma/f_0 (500)$ and $a_0 (980)$ scalars have been reconsidered  in the framework of the linearized 
Nambu--Jona-Lasinio model. In Ref. \cite{PAUK}, the contribution from the $a_0 (980)$, $f_0 (980)$, $f_0 (1370)$
states were evaluated as single-meson exchange terms with phenomenological form factors, see Fig. \ref{fig:scalar}.
Finally, the contribution from the lightest scalar, the $\sigma/f_0 (500)$ is contained
in the dispersive evaluation of the contribution to HLbL from two-pion intermediate states 
with $\pi\pi$ rescattering of Refs. \cite{Colangelo:2017qdm,Colangelo:2017fiz}.

%%%%%%%%%%%%%%%%%%%%%%%%%%%%%%%%%%%%%%%%%%%%%%

The approach considered here for the treatment of the contribution from scalar states
to HLbL has, to some extent, overlaps with both of the last two of these more recent approaches. It rests
on a set of coupled-channel dispersion relations for the processes $\gamma\gamma\to\pi\pi, K{\bar K}$,
where the strong S-matrix amplitudes for $\pi\pi\to\pi\pi, K{\bar K}$ are represented by
an analytic K-matrix model, first introduced
in Ref. \cite{MENES}, and gradually improved over time in Refs. \cite{OCHS,WANG,WANG2},
as more precise data on $\pi\pi$ scattering and on the reactions $\pi\pi\to\gamma\gamma$
became available. The details of the model will not be discussed here, as they are
amply documented in the quoted references. The interest for our present purposes of the 
analysis of the data within this K-matrix framework is twofold. First, it contributes
to our knowledge of the two-photon widths of some of the scalar states, which we will
need as input. Second, through the fit to data of the K-matrix description of $\pi\pi$ scattering,
it provides information on the mass and the total hadronic width of the $\sigma/f_0(500)$ resonance,
which will also be needed.

The rest of this article is organized as follows. Section \ref{sec:def} briefly recalls
the basic formalism describing the hadronic light-by-light contribution to the anomalous
magnetic moment of a charged lepton. This is then specialized to the contribution due to the
exchange of a narrow-width scalar state (Section \ref{sec:scalar_lbl}).
Some relevant properties of the vertex function involved are discussed in Section \ref{sec:Gamma^S}, where
a vector-meson-dominance (VMD) representation satisfying its leading short-distance behaviour is also given. Three sections are
devoted to a review of the properties (mass and width) of the $f_0/\sigma$ scalar, coming
either from sum rules (Section \ref{sec:scalars}) or from phenomenology (Section \ref{sec:sigma}).
In Section \ref{sec:sigma} we furthermore describe how our formalism also allows to handle broad resonances like
$\sigma/f_0 (500)$ or $f_0 (1370)$. The values of the mass and of the width of the $\sigma/f_0 (500)$
retained for the present study are given in the last of these three sections 
(Section \ref{sec:sigma_values}). The two-photon widths of the remaining scalar mesons
are discussed in Section \ref{sec:scalar_widths}. Our results concerning the contributions
of the scalars to HLbL are presented and discussed in Section \ref{sec:a^lbl_values}.
Finally, we summarize the present experimental and theoretical situation
concerning the standard-model evaluation of the anomalous magnetic moment of the muon
(Section \ref{sec:status}) and end this article by giving our conclusions (Section \ref{sec:concl}).

%%%%%%%%%%%%%%%%%%%%%%%%%%%%%%%%%%%%%%%
\section{Hadronic light-by-light contribution to $a_l$}
\label{sec:def} 
\setcounter{equation}{0}
%%%%%%%%%%%%%%%%%%%%%%%%%%%%%%%%%%%%%%%

The hadronic light-by-light
contribution to the muon anomalous magnetic moment, illustrated in Fig. \ref{fig:box}, is equal to \cite{Aldins:1970id}:

\bea
&a_\mu^{lbl}\equiv {F}_2(k=0)\nnb\\
&=\frac{1}{48m}{\mbox{tr}}\left\{(\not \!\!p + m)[\gamma^{\rho},\gamma^{\sigma}](\not \!\!p + m){\Gamma}_{\rho\sigma}(p,p)\right\}
\lbl{F2trace}
\eea
where $k$ is the momentum of the external photon, while $m$ and  $p$ denote the muon mass and momentum.
Furthermore [$p'=p+k$ ]
\bea
&&{%\widehat
\Gamma}_{\rho\sigma}(p\,',p)
\equiv
% \nnb\\ &&
-ie^6\hspace*{-0.15cm}\int \hspace*{-0.15cm} \frac{d^4q_1}{(2\pi)^4}\hspace*{-0.15cm}\int \hspace*{-0.15cm} \frac{d^4q_2}{(2\pi)^4}
\frac{1}{q_1^2\,q_2^2\,(q_1+q_2-k)^2}
\nonumber\\
&&\times
\frac{1}{(p\,'-q_1)^2-m^2}\,
\frac{1}{(p\,'-q_1-q_2)^2-m^2}
\nonumber\\
&&\times
\gamma^{\mu}(\not\! p\,'- \not\!q_1+m)
\gamma^{\nu}(\not\! p\,'- \not\! q_1-\not\! q_2+m)
\gamma^{\lambda}
\nonumber\\
&&\times
\frac{\partial}{\partial k^{\rho}}\,
\Pi_{\mu\nu\lambda\sigma}(q_1,q_2,k-q_1-q_2)\, , 
\lbl{Gamma2}
\eea
with $q_1, q_2, q_3$ the momenta or the virtual photons and
\bea
\Pi_{\mu\nu\lambda\rho}(q_1,q_2,q_3)\hspace*{-0.10cm} =\hspace*{-0.15cm}
%\nnb\\&&\quad
\int \hspace*{-0.15cm}d^4x_1\hspace*{-0.15cm}\int\hspace*{-0.15cm} d^4x_2\hspace*{-0.15cm}\int\hspace*{-0.15cm} d^4x_3
\,e^{i(q_1\cdot x_1 + q_2\cdot x_2 + q_3\cdot x_3)}\,
\nonumber\\
\!\!\!\times
%\quad\quad\quad
\langle\,0\,\vert\,\mbox{T}
\{j_{\mu}(x_1)j_{\nu}(x_2)j_{\lambda}(x_3)j_{\rho}(0)\}  
\,\vert\,0\,\rangle
\label{eq:4rank}
\eea
the fourth-rank light quark  vacuum polarization tensor, $j_{\mu}$ the electromagnetic 
current and $\vert\,0\,\rangle$ the QCD vacuum.
%%%%%%%%%%%%%%%%%%%%%%%%%

In practice, the computation of $a_\mu^{lbl}$ involves the limit 
$k\equiv p\,' - p \to 0$ of an expression of the type:
\bea
\vspace*{-1cm}
{\cal F}(p\,'\! ,p) \!
&=& \!- ie^6 \!\int \!\frac{d^4q_1}{(2\pi)^4} \int \!\frac{d^4q_2}{(2\pi)^4}\,
%\times\nnb\\&&\quad
{\cal J}^{\mu\nu\rho\sigma\tau}(p',p ; q_1,q_2)
\nonumber\\
&&\times{\cal F}_{\mu\nu\rho\sigma\tau}(-q_1,q_2+ q_1 + k,-q_2,-k)
,~~~
\eea
 where
%\vspace*{-0.5cm}
\par
\begin{strip}
\bea
{\cal J}^{\mu\nu\rho\sigma\tau}(p\,',p\, ; q_1,q_2) &=&
%\nnb\\ &&\quad
\frac{1}{(p\,'+q_1)^2-m^2}\,
\frac{1}{(p -q_2)^2-m^2} \,
\frac{1}{q_1^2\,q_2^2\,(q_1+q_2+k)^2}\nnb\\
&&\times
\frac{1}{48 m}\,{\mbox{tr}}[ (\not\!\! p + m) [ \gamma^\sigma ,\gamma^\tau ]
(\not\!\! p\,' + m)
\gamma^{\mu}
(\not\!\! p\,' + \not\!\! q_1+m)
\gamma^{\nu}
(\not\!\! p \, - \not\!\! q_2 + m)\gamma^\rho ] .%\nnb\\
\eea
\end{strip}
\par

\nin
This tensor has the symmetry property
%\bea\lbl{symJ}
$
{\cal J}^{\mu\nu\rho\sigma\tau}(p\,',p\, ; q_1,q_2) =$ $
{\cal J}^{\rho\nu\mu\tau\sigma}(p, p\,' ; -q_2,-q_1),$
%\eea
%\item On the other hand, 
while, due to Lorentz invariance, ${\cal F}(p\,'\! ,p)$ depends on the momenta $p$ and $p\,'$
through their invariants only. For on-shell leptons, $p^2 = p^{\,\prime 2} = m^2$, this amounts to
${\cal F}(p\,'\! ,p) \equiv {\cal F}(k^2) = {\cal F}(p,p\,')$.

 %%%%%%%%%%%%%%%%%%%%%%%%%%%%%%%%%%%%%%%
%Having previously fixed the values of the scalar mesons masses and $\gamma\gamma$ couplings, we are now ready to discuss their quantitative contributions to $a_\mu$ from LbL.
\section{Scalar meson contributions to $a_\mu^{lbl}$}
\label{sec:scalar_lbl} 
\setcounter{equation}{0}
%%%%%%%%%%%%%%%%%%%%%%%%%%%%%%%%%%%%%%%
Let us focus on the contribution to $a_\ell^{lbl}$ due to the exchange
of a $0^{++}$ scalar meson $S$. We first discuss the situation where the width of this scalar
meson is small enough so that its effects can be neglected. As a look to Table \ref{tab:PDG_scalars}
shows, this will be the case for $S = a_0 (980), f_0 (980), f_0 (1500)$. The circumstances under which 
the quite broad $\sigma/f_0(500)$ resonance, and possibly also the $f_0 (1370)$ state, can be 
treated in a similar manner will be addressed in due course. 

%%%%%%%%%%%%%%%%%%%%%%%%%%%%%%%%%%%
 {\scriptsize
\begin{table}[b]
\begin{center}
%\begin{table*}[hbt]
\setlength{\tabcolsep}{1.5pc}
 \caption{\footnotesize    
The scalar states we consider together with the estimates or averages for the mass and width, as given
by the 2018 Edition of the Review of Particle Physics \cite{RPP18}. In
the cases of the $\sigma/f_0 (500)$ and $f_0 (1370)$ states, the ranges
represent the estimates of the Breit-Wigner masses and widths.}
\vspace*{0.25cm}
    {\footnotesize
\begin{tabular}{crr}
\hline
\hline
\\
Scalar &  Mass [MeV]   & Width [MeV]
\\
\\
\hline
\\
$\sigma/f_0 (500)$ & 400 --550 &  400 -- 700
\\
$a_0 (980) $ & 980(20) & 50 -- 100
\\
$f_0 (990) $ & 990(20) & 10 -- 100
\\
$f_0 (1370) $ & 1200 -- 1500 & 200 -- 500
\\
$a_0 (1450) $ & 1474(19) &  265(13)
\\
$f_0 (1500) $ & 1504(6) & 109(7)
\\
\\
\hline\hline
\end{tabular}
}
\label{tab:PDG_scalars}
\end{center}
\end{table}
}
\nin
%%%%%%%%%%%%%%%%%%%%%%%%%%%%%%%%%%%%%%%%%%

The contribution 
$\Pi_{\mu\nu\rho\sigma}^{(S)}(q_1,q_2,q_3)$ due to the exchange of a scalar one-particle state 
$\vert S (p_S) \rangle$ to the fourth-order vacuum-polarization tensor $\Pi_{\mu\nu\rho\sigma} (q_1,q_2,q_3)$
(see Fig. \ref{fig:box}) is described  by the Feynman diagrams shown in Fig.\ref{fig:scalar}. 
It involves the form factors describing the photon-photon-scalar vertex function
\bea
\Gamma^S_{\mu\nu} (q_1 ; q_2)\hspace*{-0.2cm}& \equiv&\hspace*{-0.2cm}
%\nnb\\ &&
i \int d^4 x \, e^{-i q_1 \cdot x} 
\langle 0 \vert T \{ j_\mu (x) j_\nu (0) \} \vert S (p_S) \rangle
\nnb\\
&=&\hspace*{-0.2cm}{\cal P}(q^2_1 , q^2_2) P_{\mu\nu}(q_1 , q_2) + {\cal Q}(q_1^2 , q_2^2) Q_{\mu\nu}(q_1 , q_2).\nnb\\
\lbl{S_FF}
\eea
where  $q_2 \equiv p_S - q_1$. 
This decomposition of $\Gamma^S_{\mu\nu} (q_1 ; q_2)$
follows from Lorentz invariance, invariance under parity, and the conservation of the 
current $j_\mu (x)$. The tensors
\bea
P_{\mu\nu}(q_1 , q_2) &=& q_{1,\nu} q_{2,\mu} - \eta_{\mu\nu} (q_1\cdot q_2) ,
\nonumber\\
Q_{\mu\nu}(q_1 , q_2) &=& q_2^2 q_{1,\mu} q_{1,\nu} + q_1^2 q_{2,\mu} q_{2,\nu} \nnb\\
&&- (q_1 \cdot q_2) q_{1,\mu} q_{2,\nu}
- q_1^2 q_2^2 \eta_{\mu\nu} ,
\label{eq:invariant}
\eea
are transverse,
%\be
%\{ q_1^\mu ; q_2^\nu \} P_{\mu\nu}(q_1 , q_2) = \{ 0 \,; 0 \}
%,\quad
%\{ q_1^\mu ; q_2^\nu \} Q_{\mu\nu}(q_1 , q_2) = \{ 0 \,; 0 \},
%\ee
\be
q_{1,2}^{\mu,\nu}  P_{\mu\nu}(q_1 , q_2) =0
,\quad
q_{1,2}^{\mu,\nu} Q_{\mu\nu}(q_1 , q_2) = 0,
\ee
and symmetric under the simultaneous exchanges of the momenta $q_1$ and $q_2$
and of the Lorentz indices $\mu$ and $\nu$.
The two off-shell scalar-photon-photon transition form factors ${\cal P}(q_1 , q_2)$ and 
${\cal Q}(q_1 , q_2)$ depend only on the two independent invariants $q_1^2$ and $q_2^2$, 
and, are symmetric under permutation of the momenta $q_1$ and $q_2$. It is important
to point out that the amplitude for the decay $S\to\gamma\gamma$, which is proportional
to ${\cal P}(0,0) M_S^2 (\epsilon_1 \cdot \epsilon_2)$
[$\epsilon_i$ denote the respective photon polarization vectors, which are transverse, $q_i \cdot\epsilon_j=0$],
provides information on ${\cal P}(0,0)$ only.

In order to simplify subsequent formulas, we will use the following short-hand notation:
\be
P_{\mu\nu}(q_i , q_j) \equiv P_{\mu\nu}^{(i,j)},
\quad
Q_{\mu\nu}(q_i , q_j) \equiv Q_{\mu\nu}^{(i,j)},
%\label{eq:invariant}
\ee
and
\bea
{\cal P}(q^2_i , q^2_j)&\equiv& {\cal P}_{(i,j)}\quad; \quad {\cal P}[q_i^2 , (q_j+q_k)^2]\equiv {\cal P}_{(i,jk)}~,\nnb\\
{\cal Q}(q^2_i , q^2_j)&\equiv& {\cal Q}_{(i,j)}\quad ; \quad {\cal Q}[q_i^2 , (q_j+q_k)^2]\equiv {\cal Q}_{(i,jk)}~,\nnb\\
{\cal P}(q_i^2 , 0)&\equiv& {\cal P}_{(i,0)}\quad; \quad {\cal Q}(q_i^2 , 0)\equiv {\cal Q}_{(i,0)}~,
\label{eq:function}
\eea
The contribution $a_{\mu}^{lbl}\vert_S$ to $a_\mu^{lbl}$ from the exchange of the scalar $S$ is then obtained upon replacing,
in the general formula (\ref{eq:4rank}), the tensor $\Pi_{\mu\nu\rho\sigma}(q_1,q_2,q_3,q_4)$ by
\vspace*{-0.35cm}
\par
\begin{strip}
%%%%%%%%%%%%%%%%%%%%%%%%%%%%%%%%
%\vspace*{0.5cm}
%\begin{strip}
\bea
&& i\Pi_{\mu\nu\rho\sigma}^{(S)}(q_1,q_2,q_3,q_4)=
%\nnb\\ &&
%\frac{(-i)}{(q_1 + q_2)^2 - M_{S}^2}\times
 {\cal D}_S^{(1,2)}
\left[
{\cal P}_{(1,2)} P_{\mu\nu}^{(1,2)} + {\cal Q}_{(1,2)} Q_{\mu\nu}^{(1,2)}
\right]
%&&\quad\quad\quad\quad
%\times
\left[
{\cal P}_{(3,4)} P_{\rho\sigma}^{(3,4)} + {\cal Q}_{(3,4)} Q_{\rho\sigma}^{(3,4)}
\right]
%&&\quad\quad\quad\quad
\nonumber\\
&&\hspace*{3.2cm}
+{\cal D}_S^{(1,3)}
\left[
{\cal P}_{(1,3)} P_{\mu\rho}^{(1,3)} + {\cal Q}_{(1,3)} Q_{\mu\rho}^{(1,3)}
\right] \left[
{\cal P}_{(2,4)} P_{\nu\sigma}^{(2,4)} + {\cal Q}_{(2,4)} Q_{\nu\sigma}^{(2,4)}
\right]
%&& \quad\quad\quad\quad
\nonumber\\
&&\hspace*{3.2cm}
 +{\cal D}_S^{(1,4)}
\left[
{\cal P}_{(1,4)} P_{\mu\sigma}^{(1,4)} + {\cal Q}_{(1,4)} Q_{\mu\sigma}^{(1,4)}
\right]
%&&\quad\quad\quad\quad
%\times
\left[
{\cal P}_{(2,3)} P_{\nu\rho}^{(2,3)} + {\cal Q}_{(2,3)} Q_{\nu\rho}^{(2,3)}
\right]
\nnb\\
%\quad\quad\quad\quad
&&\hspace*{3.0cm}\equiv 
i \Big{\{}\Pi_{\mu\nu\rho\sigma}^{(S;PP)}%(q_1,q_2,q_3) 
+
\Pi_{\mu\nu\rho\sigma}^{(S;PQ)}%(q_1,q_2,q_3) 
+
\Pi_{\mu\nu\rho\sigma}^{(S;QQ)}%(q_1,q_2,q_3)
\Big{\}}, 
%\qquad ~ 
\lbl{Pi_S}
\eea
\end{strip}
\par
%\end{strip}
%%%%%%%%%%%%%%%%%%%%%%%%%%%
\noindent
where  $q_4^\mu \equiv - (q_1 + q_2 + q_3)^\mu$. 
The scalar-meson propagator  in the Narrow Width Approximation (NWA) reads
\beq
{\cal D}_S^{(i)}\equiv \frac{1}{q_i^2 - M_{S}^2}~ ; \quad
{\cal D}_S^{(i,j)}\equiv \frac{1}{(q_i + q_j)^2 - M_{S}^2}~,
\eeq
with $i,j=1,..4$.
In the last line, the first (third) term collects all the contributions quadratic
in the form factor ${\cal P}$ (${\cal Q}$), while the second term collects all
the contributions involving the products ${\cal P}{\cal Q}$ of the two kinds of
form factors. Correspondingly, we perform the decomposition
$a_{\mu}^{lbl}\vert_S = a_{\mu}^{lbl}\vert_S^{PP} + a_{\mu}^{lbl}\vert_S^{PQ} + a_{\mu}^{lbl}\vert_S^{QQ}$.

Starting from the representation \rf{Pi_S}, it is a straightforward
exercise to insert it into the general expression in Eq. (\ref{eq:4rank}), and then to
compute the projection on the Pauli form factor as defined 
in Eq. \rf{F2trace}. 
For further use, we introduce the tensor
%\beq
${\cal F}_{\mu\alpha\beta} (q) = \eta_{\mu\beta}q_\alpha -  \eta_{\mu\alpha}q_\beta$,
%\eeq
%the two photon form factor using a Vector Meson Dominance parametrization:
%\bea
%{\cal P}(Q_1,Q_2)&=&\tilde g_{S\gamma\gamma}{M_V^2\over Q_1^2+M_V^2}{M_V^2\over Q_2^2+M_V^2},
%\eea
and the amplitude
\be
A_{S}^{PP} (q_1 , q_2 , q_3, q_4) \equiv {\cal D}^{(1,2)}_S
{\cal P}_{(1,2)} {\cal P}_{(3,4)},
\ee
and similarly for other products of form factors $PQ,QQ$. \\
The part of the scalar-exchange term that involves the form factor
${\cal P}$ alone then reads
\bea
 a_{\mu}^{lbl}\vert_S^{PP}\ \hspace*{-0.20cm}&=&\hspace*{-0.20cm}
- e^6\,\int\frac{d^4q_1}{(2\pi)^4}\int\frac{d^4q_2}{(2\pi)^4}
%\nnb\\&\times&\quad
{\cal J}^{\mu\nu\rho\sigma\tau}(p,p\, ; q_1,q_2)
\nonumber\\
&&\hspace*{-0.25cm}\times
\,\bigg\{
2 A_{S} ^{PP}(-q_1,q_1 + q_2,-q_2,0)
%\nnb\\&&\times
{\cal F}_{\mu\nu\alpha}(q_1) (q_1 +q_2)^\alpha \nnb\\
&&\hspace*{-0.25cm} \times\ {\cal F}_{\rho\sigma\tau}(q_2)
%\nonumber\\&&
+
%\quad
 A_{S} ^{PP}(-q_1,-q_2,q_1 + q_2,0)\nnb\\
&&\hspace*{-0.25cm}\times\ {\cal F}_{\mu\rho\alpha}(q_1) q_2^\alpha  {\cal F}_{\nu\sigma\tau}(q_1+q_2)
\bigg\},
\label{eq:t1}
\eea
where the symmetry properties of the integrand, and of the amplitude
$A_{S} (q_1,q_2,q_3,q_4)$, as well as 
${\cal F}_{\rho\sigma\tau}(q) = -{\cal F}_{\rho\tau\sigma}(q)$ have been used. 
%%%%%%%%%%%%%%%%%%%%%%%%%%%%%%%%%%%%%%%
Noticing that $Q_{\mu\nu}(q , k)$ is quadratic in the components
of the momentum $k^\mu$, one sees that all of $\Pi_{\mu\nu\rho\sigma}^{(S;QQ)}(q_1,q_2,q_3)$
and half of the terms in $\Pi_{\mu\nu\rho\sigma}^{(S;PQ)}(q_1,q_2,q_3)$
will not contribute to the Pauli form factor at vanishing momentum transfer.
The part of the scalar-exchange term that involves both form factors
${\cal P}$ and ${\cal Q}$ thus reduces to
\bea
a_{\mu}^{lbl}\vert_S^{PQ}  &=&
- e^6\,\int\frac{d^4q_1}{(2\pi)^4}\int\frac{d^4q_2}{(2\pi)^4}
%\nnb\\&\times&
{\cal J}^{\mu\nu\rho\sigma\tau}(p,p\, ; q_1,q_2)\nnb\\
&&\times\,\bigg\{  
2 A^{PQ}_S(-q_2, 0 , -q_1,q_1 + q_2 )
%\nnb\\&&\times\quad
{\mathcal F}_{\rho\sigma\tau}(-q_2) \nnb\\
&&\times\ Q_{\mu\nu}(q_1, q_1 + q_2 )
 +\,
A^{PQ}_S (q_1 + q_2 , 0 , q_1, q_2) 
\nnb\\
&&\times\ {\mathcal F}_{\nu\sigma\tau}(q_1 + q_2) Q_{\mu\rho}(q_1,q_2)
\bigg\} ,
\eea
whereas $a_{\mu}^{lbl}\vert_S^{QQ}=0$.
%%%%%%%%%%%%%%%%%%%%%%%%%%%%%%%%%%%%%%%
%
The trace calculation\footnote{The corresponding Dirac traces have been
computed using the FeynCalc package~\cite{Mertig:1990an,Shtabovenko:2016sxi}.} leads to the final expression
\bea
&&\hspace{-1.3cm} 
a_\mu^{lbl}\vert_S =
-  e^6\int\frac{d^4q_1}{(2\pi)^4}\int\frac{d^4q_2}{(2\pi)^4}\times\,\nnb\\
&&\hspace{-0.8cm}
\frac{1}{q_1^2\,q_2^2\,(q_1+q_2)^2}
%\nnb\\
%\nonumber\\
%&&\quad
%\times
%&&\times
\frac{1}{(p + q_1)^2-m^2}\,
\frac{1}{(p - q_2)^2-m^2}
\nonumber\\
&&\hspace{-0.8cm}%\!\!\!\!\!\!\!\!\!\!
\Big\{
{\cal D}^{(2)}_S\big{[}{\cal P}_{(1,12)} {\cal P}_{(2)}\,  T_{1,S}^{PP} 
+ {\cal P}_{(2)} {\cal Q}_{(1,12)}\,  T_{1,S}^{PQ} \big{]}
%\right.
\nonumber\\
&&\hspace{-1.0cm}
%\left.
 \,+\,
{\cal D}^{(1,2)}_S\big{[}{\cal P}_{(1,2)} {\cal P}_{(12, 0)} \,  T_{2,S}^{PP}
+{\cal P}_{(12,0)} {\cal Q}_{(12, 0)} \,  T_{2,S}^{PQ}\big{]}
\Big\},%\nnb\\
%\hspace{1.2cm}
\label{eq:amu-mink}
\eea
where the amplitudes $T_{i,S}$ are given in Table \ref{tab:ampli} and the functions
${\cal P}_{(i,j)}$ and ${\cal Q}_{(i,j)}$ in Eq. (\ref{eq:function}). 
%%%%%%%%%%%%%%%%%%%%%%%%%%%%%%%%%%%%%%%%%%
\begin{table*}[hbt]
\setlength{\tabcolsep}{1.pc}
\caption{Expressions, in Minkowski space, of the amplitudes  defined in Eq. (\ref{eq:amu-mink}).}
\begin{tabular*}{\textwidth}{@{}l@{\extracolsep{\fill}}l}
&\\
\hline
&\\
$T_{1,S}^{PP}(q_1,q_2)$ =&
$\frac{16}{3}\Big{[}
q_2^2(p\cdot q_1)^2 + q_1^2(p\cdot q_2)^2 - (q_1\cdot q_2)(p\cdot q_1) (p\cdot q_2)
 + \,
(p\cdot q_1) (p\cdot q_2) q_2^2 + (p\cdot q_1) (q_1\cdot q_2) q_2^2 

-$\\
&$  (p\cdot q_2) (q_1\cdot q_2)^2 - m_{\ell}^2 q_1^2 q_2^2 - m_{\ell}^2 (q_1\cdot q_2) q_2^2
\Big{] }

 + \, 8 (p\cdot q_1) q_1^2 q_2^2 - 8 q_1^2 (p\cdot q_2) (q_1\cdot q_2),$ \\
 
$T_{2,S}^{PP}(q_1,q_2)=$ &$
\frac{8}{3}\Big{[}
q_2^2(p\cdot q_1)^2 + q_1^2(p\cdot q_2)^2- 2 (q_1 + q_2)^2 (p\cdot q_1) (p\cdot q_2)
+\, (p\cdot q_1) (p\cdot q_2) q_1^2 + (p\cdot q_1) (p\cdot q_2) q_2^2+$\\
&$ m^2 (q_1\cdot q_2) (q_1 + q_2)^2
\Big{]} .$\\

$T_{1,S}^{PQ}(q_1,q_2) =$&
$\frac{16}{3}\Big{[}
(q_1\cdot q_2)(p\cdot q_1) (p\cdot q_2) (q_1^2 + q_2^2) +\, (p\cdot q_1) (p\cdot q_2) (q_1\cdot q_2)^2
+(p\cdot q_2)(q_1\cdot q_2) q_1^2 q_2^2
-\, q_1^2 q_2^2(p\cdot q_1)^2 -$\\
&$ q_1^2 q_2^2 (p\cdot q_2)^2 -\, q_2^2 (q_1\cdot q_2) (p\cdot q_1)^2 -q_1^2 (q_1\cdot q_2) (p\cdot q_2)^2-\, 
(p\cdot q_1)^2 q_2^4 - (p\cdot q_2)^2 q_1^4- (p\cdot q_1) q_1^2 q_2^4  -$\\
&$(p\cdot q_1) (p\cdot q_2) q_1^2 q_2^2
+ m_{\ell}^2 q_1^2 q_2^2
(q_1 + q_2)^2\big{] }-\frac{40}{3} \big{[} (p\cdot q_1) (q_1\cdot q_2) q_1^2 q_2^2 - (p\cdot q_2) (q_1\cdot q_2)^2 q_1^2 \Big{]}$\\

&$+ 8 q_1^4 \big{[}(p \cdot q_2) (q_1 \cdot q_2) -  q_2^2 (p \cdot q_1) \big{]},$\\

$T_{2,S}^{PQ}(q_1,q_2) =$&
$\frac{4}{3}\Big{[}
2(q_1\cdot q_2)(p\cdot q_1) (p\cdot q_2) (q_1^2 + q_2^2) 
 -  q_2^2 (q_1 + q_2)^2(p\cdot q_1)^2-q_1^2 (q_1 + q_2)^2 (p\cdot q_2)^2
  -\, q_2^2 (q_1^2 + q_2^2) (p\cdot q_1)^2 $\\
&$
-  q_1^2 (q_1^2 + q_2^2) (p\cdot q_2)^2+ 2 m_{\ell}^2 q_1^2 q_2^2 (q_1 + q_2)^2
\Big{]}.$\\
\\
\hline
\end{tabular*}
\label{tab:ampli}
\end{table*}
\nin
%%%%%%%%%%%%%%%%%%%%%%%%%%%%%%%%%%%%%%%%%%
 Let us simply note here that $T_1^{(PP)}(q_1,q_2)$ and $T_1^{(PQ)}(q_1,q_2)$
come from the sum of the two diagrams $(a)$ and $(b)$ of Fig. \ref{fig:scalar} (they give identical
contributions), while $T_2^{(PP)}(q_1,q_2)$ and $T_2^{(PQ)}(q_1,q_2)$ represent the contributions
from diagram $(c)$. Apart from the presence of two form factors, the situation, at this level, is
similar to the one encountered in the case of the exchange of a pseudoscalar meson,
see for instance  Ref. \cite{KN}.

%%%%%%%%%%%%%%%%%%%%%%%%%%%%%%%%%%%%
\section{$\Gamma^S_{\mu\nu}$ %(q_1,q_2)$ 
at short distance and Vector Meson Dominance}
\label{sec:Gamma^S} 
\setcounter{equation}{0}
%%%%%%%%%%%%%%%%%%%%%%%%%%%%%%%%%%%%%

In order to proceed, some information about the vertex function
$\Gamma^S_{\mu\nu} (q , p_S - q)$ is required. In particular, the question
about the relative sizes of the contributions to %$a_\ell^{{\mbox{\tiny{LxL;$S$}}}}$ 
$a_\mu^{lbl}\vert_S$ coming from the two form factors involved 
in the description of the matrix element \rf{S_FF} needs to be answered. 
In order to briefly address this issue, one first notices that at short distances
the vertex function $\Gamma^S_{\mu\nu} (q , p_S - q)$ has the following behaviour
(in the present discussion $q^\mu$ is a spacelike momentum):
\be
%&&
%\vspace*{-0.2cm}
\hspace*{-0.20cm}
\lim_{\lambda \to \infty} \Gamma^S_{\mu\nu} (\lambda q , p_S - \lambda q)
%\nnb\\&&
=
\frac{1}{\lambda^2} \left( \frac{1}{q^2} \right)^2 \Gamma^{S ; {\infty}}_{\mu\nu} (q , p_S)
+ {\mathcal O} \left( \frac{1}{\lambda^2} \right)^3 \!
,
\ee
%\vspace*{-0.2cm}
with
\bea
\lbl{asymp}
\Gamma^{S ; {\infty}}_{\mu\nu} (q , p_S) \hspace*{-0.16cm}
&=&\hspace*{-0.16cm} (q_\mu q_\nu - q^2 \eta_{\mu\nu}) A +\big[ (q \cdot p_S) q_\mu p_{S,\nu}
\\
&&\hspace{-0.56cm}
- q^2 p_{S,\mu} p_{S,\nu} + (q \cdot p_S) (q_\nu p_{S,\mu} - q \cdot p_S \eta_{\mu\nu} ) \big] B
.~
\nnb
\eea
The structure of $\Gamma^{S ; {\infty}}_{\mu\nu} (q , p_S)$ follows from the requirements
$q^\mu \Gamma^{S ; {\infty}}_{\mu\nu} (q , p_S) = 0$, $q^\nu \Gamma^{S ; {\infty}}_{\mu\nu} (q , p_S) = 0$,
and the coefficients $A$ and $B$ are combinations of the four independent ``decay constants'' which describe
the matrix elements
\bea
&&\hspace*{-0.60cm}
\langle 0 \vert : \! D_\rho {\bar\psi} Q^2 \gamma_\sigma \psi \! : \! (0) \vert S(p_S) \rangle,\, 
%\nnb\\&&
\langle 0 \vert : \! {\bar\psi} Q^2 {\mathcal M} \psi \! : \! (0) \vert S(p_S) \rangle, ~~~
\nnb\\&& \hspace*{-0.60cm}
\langle 0 \vert : \! G_{\mu\nu}^a G_{\rho\sigma}^a \! : \! (0) \vert S(p_S) \rangle,
\lbl{mathix_elements}
\eea
of the three gauge invariant local operators of dimension four that can couple to the
$0^{++}$ scalar states. Here $Q = {\rm diag} (2/3 , -1/3 , -1/3)$ denotes the
charge matrix of the light quarks, whereas ${\mathcal M} = {\rm diag} (m_u , m_d , m_s)$
stands for their mass matrix. The third matrix element, involving the gluonic operator
$: \! G_{\mu\nu}^a G_{\rho\sigma}^a \! :$,
only occurs to the extent that the scalar state possesses a singlet component. For a
pure octet state, and in the chiral limit, only one ``decay constant'', coming
from the first operator, remains, and one has $A/B = - M_S^2/2$. 
The asymptotic behaviour in Eq. \rf{asymp} leads to the suppression of ${\mathcal Q}(q_1 , q_2)$ 
with respect to ${\mathcal P}(q_1 , q_2)$ at high (space-like) photon virtualities ($Q_i^2 = - q_i^2$): 
\be
{\mathcal Q}(q_1 , q_2) \simeq - \frac{2 {\mathcal P}(q_1 , q_2)}{Q_1^2 + Q_2^2}.
\ee
This short-distance behaviour can be reproduced by a simple vector meson dominance (VMD)-type representation,
\bea
{\mathcal P}^{\rm VMD} (q_1 , q_2) &=& - \frac{1}{2} \frac{B (q_1^2 + q_2^2) +  (2A + M_S^2 B)}{(q_1^2 - M_V^2) (q_2^2 - M_V^2)}
,\nnb\\
{\mathcal Q}^{\rm VMD} (q_1 , q_2) &=& -  \frac{B}{(q_1^2 - M_V^2) (q_2^2 - M_V^2)}
,
\label{eq:formfactor}
\eea
which leads to:
\be
\kappa_S \equiv - \frac{M_S^2 {\mathcal Q}^{\rm VMD} (0 , 0)}{{\mathcal P}^{\rm VMD} (0 , 0)} = - \frac{2B M_S^2}{B M_S^2 + 2A}.
\label{kappa_def}
%\lbl{kappa_def}
\ee
Incidentally, similar statements can also be inferred from Ref. \cite{MS}, where
the octet vector-vector-scalar three-point function $\langle VVS \rangle$ was studied in the chiral
limit. From the expressions given there, one obtains
\bea
&&\frac{M_S^2{\mathcal Q}(q_1 , q_2)}{{\mathcal P}(q_1 , q_2)}
=
- \bigg[   
\frac{9}{5} \frac{M_V^4}{F_\pi^2 (M_K^2 - M_\pi^2)} {\tilde c} 
- \frac{1}{2} +\nnb\\
&& \frac{Q_1^2 + Q_2^2}{2 M_S^2} 
\bigg]^{-1} \simeq
- \frac{2 M_S^2}{2 M_S^2 + Q_1^2 + Q_2^2},
\eea
with \cite{KMS}
\bea
{\tilde c} &=& \frac{5}{16 \pi \alpha^2} \left[ \frac{\Gamma_{\rho\to e^+ e^-}}{M_\rho} 
- 3 \frac{\Gamma_{\omega\to e^+ e^-}}{M_\omega} - 3 \frac{\Gamma_{\phi\to e^+ e^-}}{M_\phi}\right]\nnb\\
&\simeq& ( 4.6 \pm 0.8 ) \cdot 10^{-3}.
\eea
Numerically, this would correspond to $A/B = - 2 M_S^2$ ($\kappa_S = 1$), rather than to $A/B = - M_S^2/2$, which, as mentioned above, 
should hold precisely for the conditions under which the analysis carried out in Ref. \cite{MS} is valid. 
This discrepancy illustrates the well-known \cite{Bijnens:2003rc,Knecht:2001xc} limitation of the simple saturation by a 
single resonance in each channel, which in general cannot simultaneously accommodate the correct short-distance behaviour 
of a given correlator and of the various related vertex functions. Let us also point out that $A/B=-M_S^2/2$ corresponds to $\mathcal P(0, 0) = 0$, i.e. to a vanishing two-photon
width. This either means that scalars without a singlet component decay into two photons through quark-mass
and/or through isospin-violating effects, or, more likely, shows the limitation of the VMD picture, which provides,
in this case, a too simplistic description of a more involved situation. The second alternative would then require to go
beyond a single-resonance description, as described, for instance, in Ref.\,\cite{Knecht:2001xc} for the photon-transition form factor
of the pseudoscalar mesons. Following this path would, however, lead us too far astray, and in the present study
we will keep the discussion within the framework set by the VMD description of the two form factors $\mathcal P(q_1, q_2)$ and $\mathcal Q(q_1, q_2)$.
For later use, like for instance the derivation of Eq.\,\ref{a_mu_integrals} below, it is also of interest to parameterize
the VMD form factors directly in terms of $\mathcal P(0, 0)$, which gives the two-photon width, and the parameter $\kappa_S$ as defined by the first equality in Eq.\,\ref{kappa_def}:
\bea
\mathcal P^{VMD}(q_1, q_2)&=&\mathcal P(0,0)\left[1-\frac{\kappa_S}{2}\frac{q_{1}^{2}+q_{2}^{2}}{M_{S}^{2}}\right]\nnb\\
&&\times \frac{M_{V}^{4}}{(q_{1}^{2}-M_{V}^{2})(q_{2}^{2}-M_{V}^{2})},\\
\mathcal Q^{VMD}(q_1, q_2)&=&-\kappa_S\frac{\mathcal P(0,0)}{M_{S}^{2}}\frac{M_{V}^{4}}{(q_{1}^{2}-M_{V}^{2})(q_{2}^{2}-M_{V}^{2})}.\nnb
\eea
We may draw two conclusions from the preceding analysis.
First, that a sensible comparison to be made, for space-like
photon virtualities, is thus not between 
${\mathcal P}(q_1 , q_2)$ and ${\mathcal Q}(q_1 , q_2)$, but rather between ${\mathcal P}(q_1 , q_2)$ and, say,
$- (2 M_S^2 + Q_1^2 + Q_2^2) {\mathcal Q}(q_1 , q_2)/2$. At high photon
virtualities, their ratio %${\widehat{\mathcal Q}}(q_1 , q_2)/{\mathcal P}(q_1 , q_2)$ 
tends to unity. Second, that $\vert {\mathcal P}(0 , 0)\vert$
and $M_S^2 \vert {\mathcal Q}(0 , 0) \vert$ may well be of comparable sizes.
For instance, within VMD, we obtain
\beq
{\mathcal P}(0 , 0) = -M_S^2 {\mathcal Q}(0 , 0) 
\lbl{eq:vmd}
\eeq
from the analysis of Ref. \cite{MS},

%%
%while from the asymptotic behaviour of the vertex function
%discussed here, we infer, under the same conditions,
%\beq
%{\mathcal P}(0 , 0) = \frac{M_S^2}{2} {\mathcal Q}(0 , 0) .
%\lbl{eq:asymptotic}
%\eeq
%Since the vertex function \rf{S_FF} is the quantity
%of interest here, we will retain the second option.
%Furthermore, since quark-mass effects are expected to be small, the 
%value $\kappa_S =-2$ is favoured for the two isovector
%states $a_0(980)$ and $a_0(1450)$ from the present
%analysis. 

%%%%%%%%%%%%%%%%%%%%%%%%%%%%%%%%%%%%%%
\section{Angular integrals}
\label{sec:ang_int} 
\setcounter{equation}{0}
%%%%%%%%%%%%%%%%%%%%%%%%%%%%%%%%%%%%%%

The next step consists in transforming the two-loop integral in Eq. (\ref{eq:amu-mink})
into an integration in Euclidian space through the replacement
\be
\int d^4 q_i \longrightarrow i(2\pi^2)\int_0^\infty \hspace*{-0.25cm}dQ_i\,Q_i^3\int \frac{d \Omega_{{\hat Q}_i}}{2\pi^2},
\ee
with $Q_i^2 = - q_i^2$, $i=1,2$, and 
\bea
\hspace*{-0.5cm} d \Omega_{{\hat Q}_i} \!\! &=& \!\! 
d \phi_{{\hat Q}_i} d \theta_{1{\hat Q}_i} d \theta_{2{\hat Q}_i} 
\sin(\theta_{1{\hat Q}_i}) \sin^2(\theta_{2{\hat Q}_i}),
\nonumber\\
 \int d \Omega_{{\hat Q}_i} \!\! &=& \!\! 2 \pi^2,
\eea
where the orientation of the four-vector $Q_\mu$ in four-dimensional Euclidian space is given by the azymuthal 
angle $\phi_{{\hat Q}}$ and the two polar angles $\theta_{1{\hat Q}}$ and $\theta_{2{\hat Q}}$.
Since the anomalous magnetic moment is a Lorentz invariant, its value does not depend on the
lepton's four-momentum $p^\mu$ beyond its mass-shell condition $p^2 = m^2$. 
One may thus average, in Euclidian space, over the directions of the four-vector $P$ (the Euclidian
counterpart of $p$, i.e. $P^2 = - m^2$)
\be
a_\mu^{lbl}\vert_S = \frac{1}{2 \pi^2} \int d \Omega_{\hat P} \, a_\mu^{lbl}\vert_S.
\ee  
This allows to obtain a representation of $a_\mu^{lbl}\vert_S^{PP+PQ}$
as an integral over three variables, $Q_1$, $Q_2$, and the angle between the two Euclidian
loop momenta \cite{JN}. Actually, in the VMD representation of Eq. (\ref{eq:formfactor}), the form 
factors belong to the general class discussed in Ref. \cite{KN}, for which one can actually 
perform the angular integrals directly, without having to average over the direction of the 
lepton four-momentum first. Within this VMD approximation of the form factors,
the anomalous magnetic moment then reads 

\vfill\newpage

\par
%Then, for the LMD form factors given in Eqs. \rf{calP_LMD} and \rf{calQ_LMD}, one
%obtains
%%%%%%%%%%%%%%%%%%
%%%%%%%%%%%%%%%%%%
%\newpage
%\vspace*{-0.35cm}
\begin{strip}
%{\foot
\bea
a_\mu^{lbl}\vert_S^{\rm VMD} \!\!\! &=& \!\!\!
 \left( \frac{\alpha}{\pi} \right)^3  [{\cal P} (0 , 0)]^2  \int_0^\infty \hspace*{-0.25cm} d Q_1 \int_0^\infty \hspace*{-0.25cm} d Q_2 
\, \frac{M_V^4}{(Q_1^2 + M_V^2) (Q_2^2 + M_V^2)} \Bigg\{ 
\Big[ \Delta w_1^{PP}(M_V) - \Delta w_2^{PP}(M_V) \Big] \nnb\\
&&
+\frac{\kappa_S}{2}
\bigg\{ \frac{Q_1^2 + Q_2^2}{M_S^2}\big[\Delta w_1^{PP}(M_V) - \Delta w_2^{PP}(M_V) \big]
- \frac{M_V^2}{M_S^2} \big[ w_{12}^{PP} (M_V)  
-2  \Delta w_1^{PQ}(M_V) -2 \Delta w_2^{PQ}(M_V) \big]\bigg\} \nnb\\
&&
+\frac{\kappa^2_S}{4}
\bigg[ \frac{Q_1^2 Q_2^2}{M_S^4} \Delta w_1^{PP}(M_V)-\tilde w_{12}^{PP}(M_V)- 2 \, \frac{Q_2^2}{M_S^2} \, \Delta w_1^{PQ}(M_V) - 2 
\, \frac{Q_1^2 + Q_2^2}{M_S^2} \, \Delta w_2^{PQ}(M_V)\bigg]\Bigg\}\nnb\\
&&\equiv
\left( \frac{\alpha}{\pi} \right)^3 [{\cal P} (0 , 0)]^2 \Big{\{} {\mathcal I}_{p} + \kappa_S {\mathcal I}_{pq} + \kappa_S^2 {\mathcal I}_{q}\Big{\}} ,
%~ {\rm from~Eq.}\,\ref{eq:vmd}\nnb\\
%&&\equiv \left( \frac{\alpha}{\pi} \right)^3 [{\cal P} (0 , 0)]^2 \Big{\{} {\mathcal I}_{p} -2{\mathcal I}_{pq}+4{\mathcal I}_{q}\Big{\}}
%~{\rm from~ Eq.}\, \ref{eq:asymptotic}\nnb\\
\label{a_mu_integrals}\lbl{a_mu_integrals}
\eea
%}
\end{strip}
\noindent
where $\kappa_S$ was defined in Eq. \ref{kappa_def}, and
with 
\be
\Delta w_{1,2}^{PP,PQ}(M) \equiv w_{1,2}^{PP,PQ} (M) - w_{1,2}^{PP,PQ} (0)
,
\ee
\bea
\lbl{eq:w12}
 %
% \Delta w_{1,2}^{PP,PQ}(M) \!\!\!&=&\!\!\! w_{1,2}^{PP,PQ} (M) - w_{1,2}^{PP,PQ} (0)
% \nnb\\
w_{12}^{PP}(M_V) \!\!\!&=&\!\!\! w_1^{PP} (M_V) -w_2^{PP} (M_V) ,\\
\tilde w_{12}^{PP}(M_V) \!\!\!&=&\!\!\! \frac{Q_2^2 M_V^2}{M_S^4} w_1^{PP} (M_V)-\frac{Q_1^2 + Q_2^2}{M_S^2} \, \frac{M_V^2}{M_S^2} \, w_2^{PP} (M_V).~
\nnb
\eea
%$ \Delta w_{1,2}^{PP,PQ}(M) \equiv w_{1,2}^{PP,PQ} (M) - w_{1,2}^{PP,PQ} (0)$.

%%%%%%%%%%%%%%%%%%%%%%%%%
\begin{figure}[H]
\begin{center}
%{\bf a)}\hspace*{-2cm} {\bf b)}\hspace*{0cm}
{\includegraphics[width=3.5cm  ]{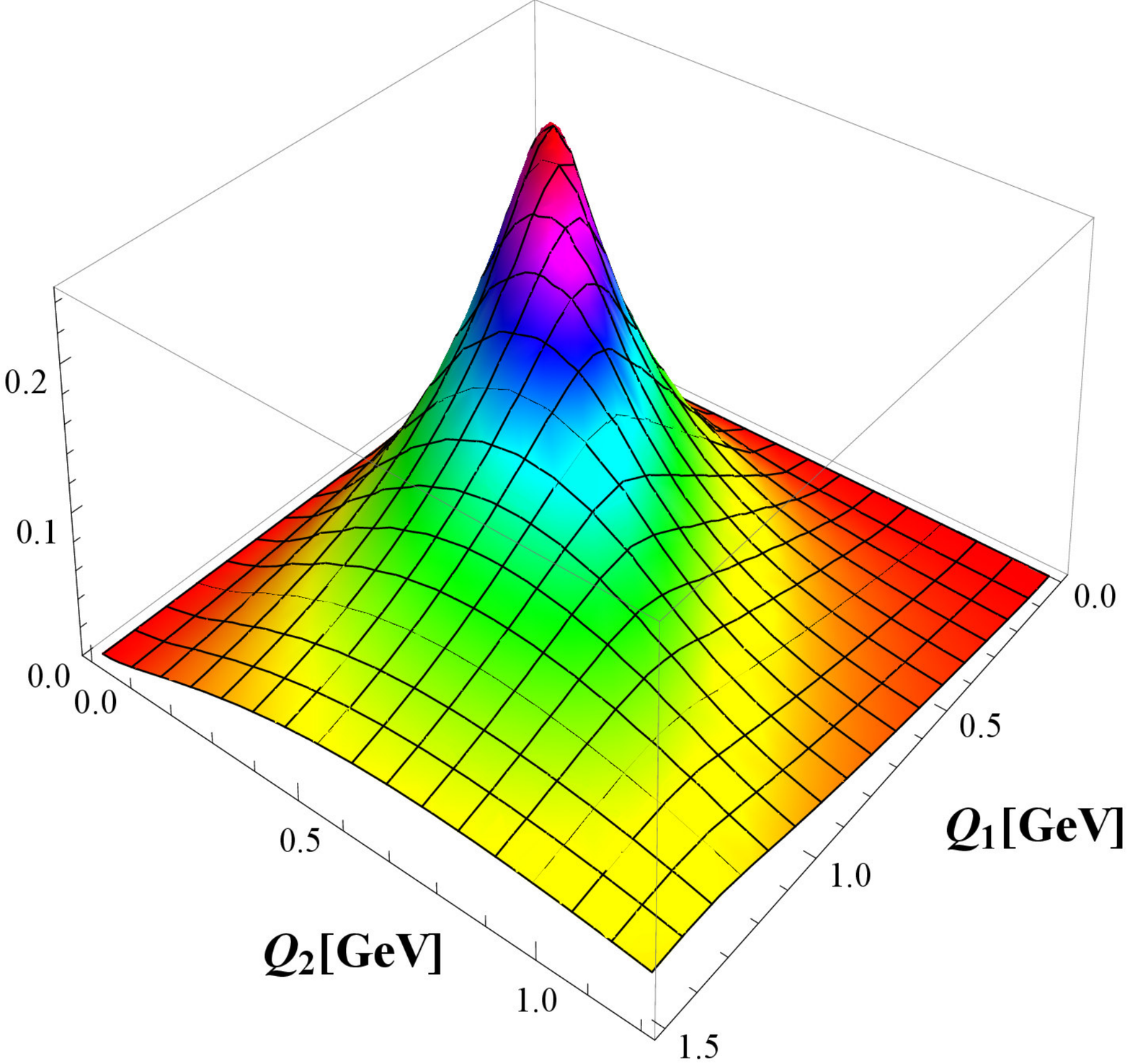}}
%\vspace*{1cm}
{\includegraphics[width=3.5cm  ]{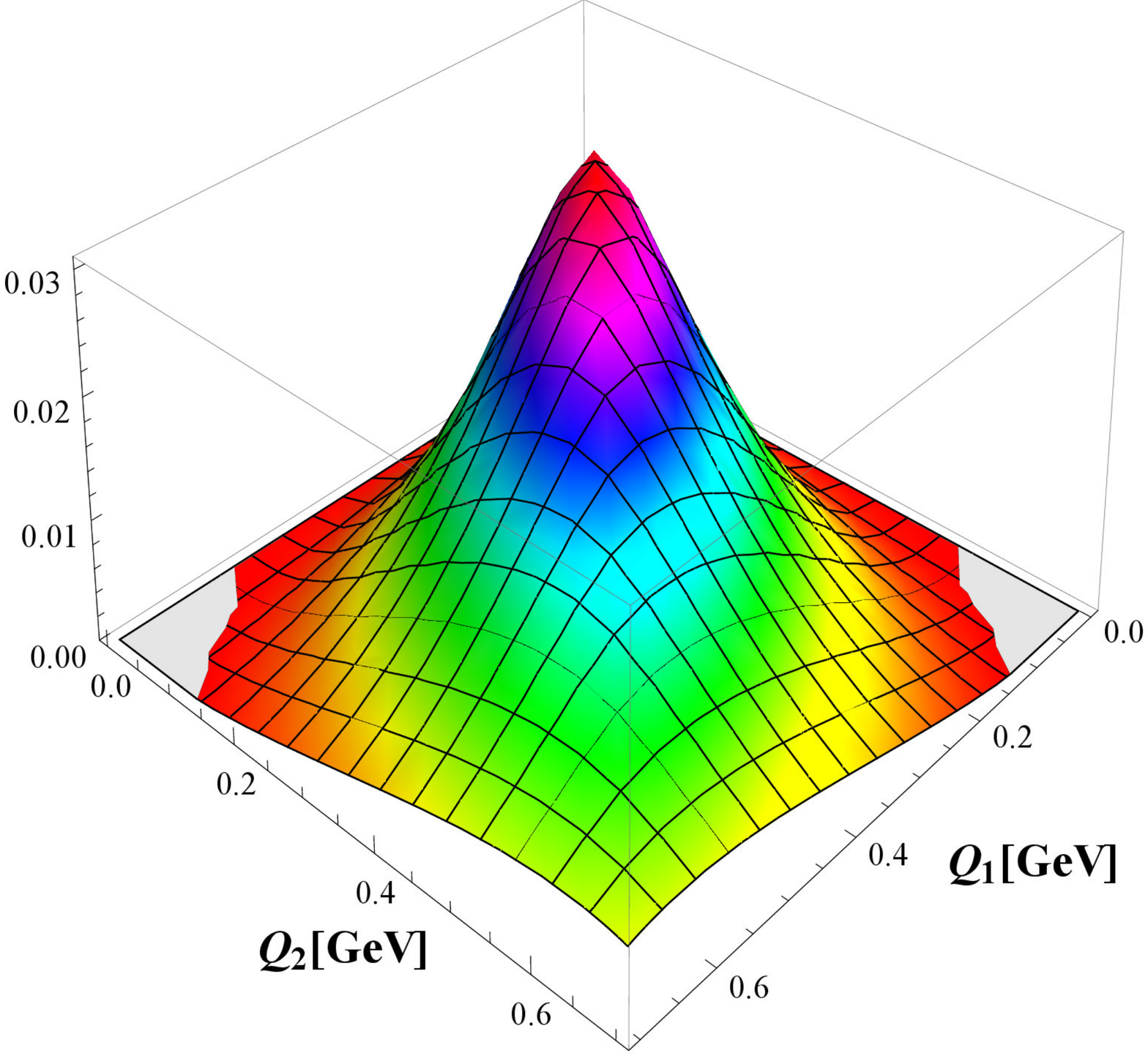}}\\
{\bf a)}\hspace*{5cm}  {\bf b)}
\caption{The weight functions: {\bf a)}: $\Delta w_1^{PP}$ and  {\bf b)}: $\Delta w_2^{PP}$ as function of $Q_1$ and $Q_2$. 
We have used $M_V=M_{\rho}=775$ MeV and $M_S=960$ MeV.}
\label{fig:weight1}
\end{center}
\end{figure}
%%%%%%%%%%%%%%%%%%%%%%%%%
\begin{figure}[H]
\begin{center}
{\includegraphics[width=3.5cm  ]{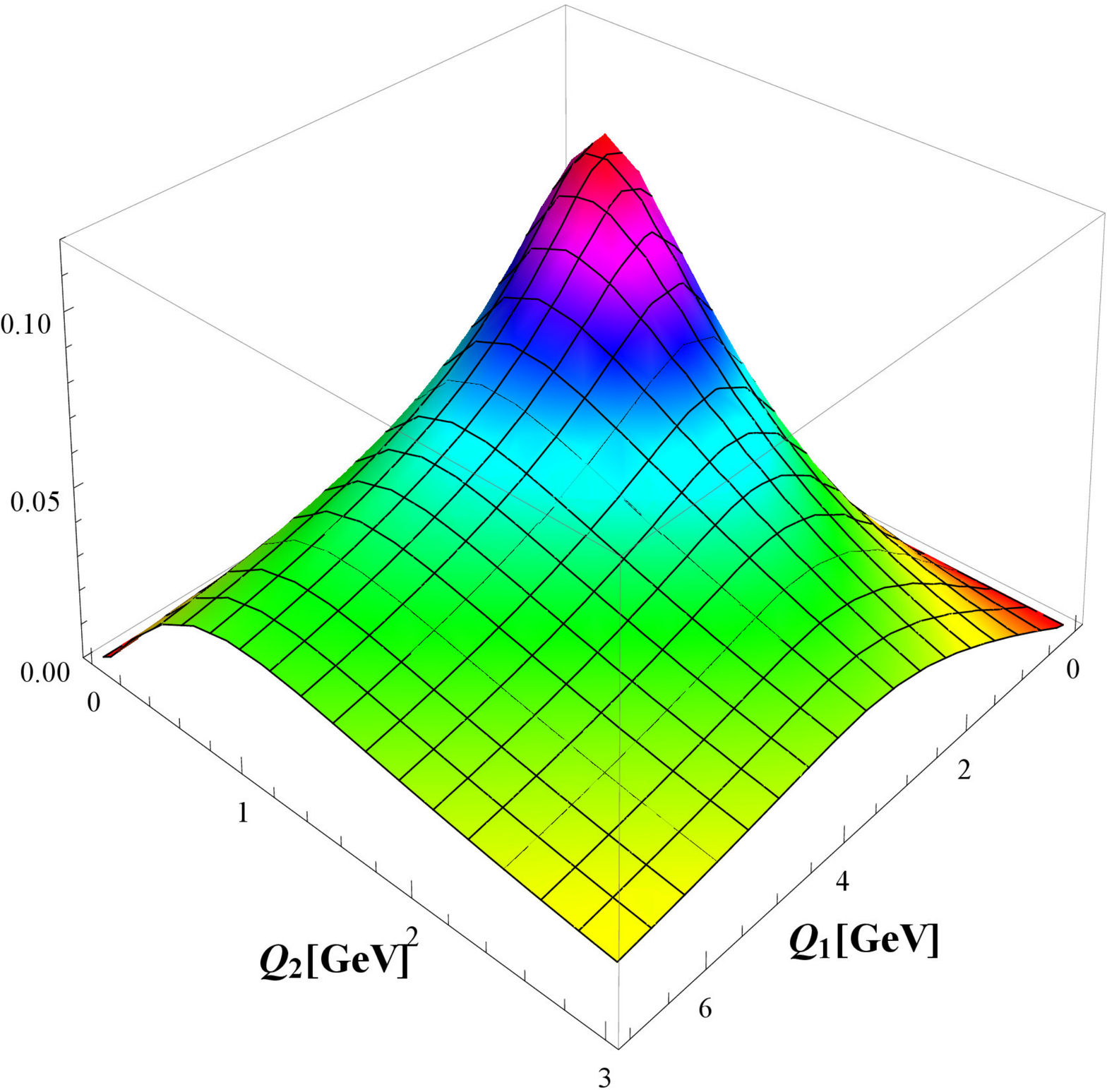}}
{\includegraphics[width=3.5cm  ]{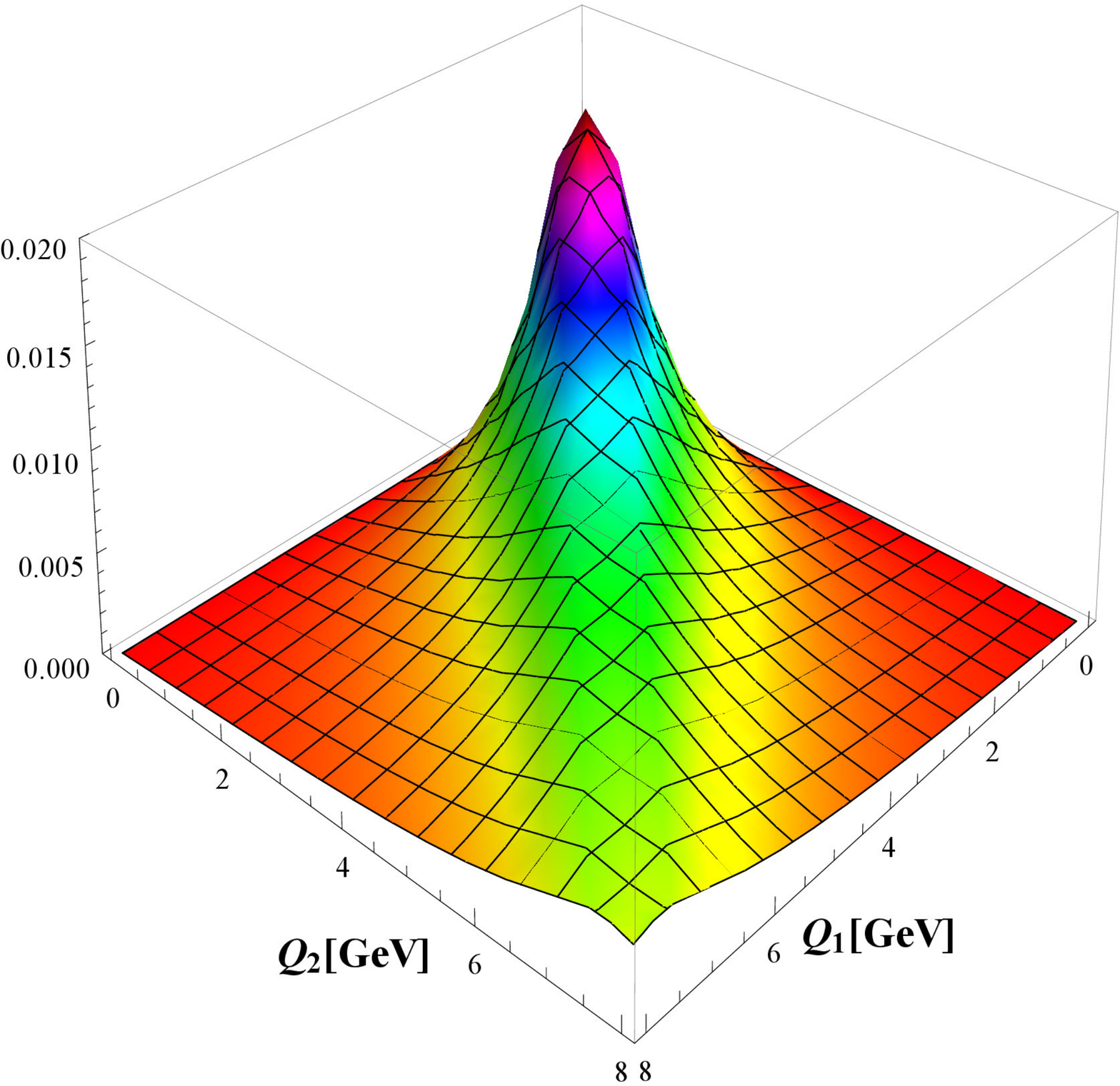}}
{\bf a)}\hspace*{5cm}  {\bf b)}
\caption{The same as in Fig. \ref{fig:weight1} but for PQ.}
\label{fig:weight2}
\end{center}
\end{figure}
%%%%%%%%%%%%%%%%%%%%%%%%%
\begin{figure}[H]
\begin{center}
{\includegraphics[width=3.5cm  ]{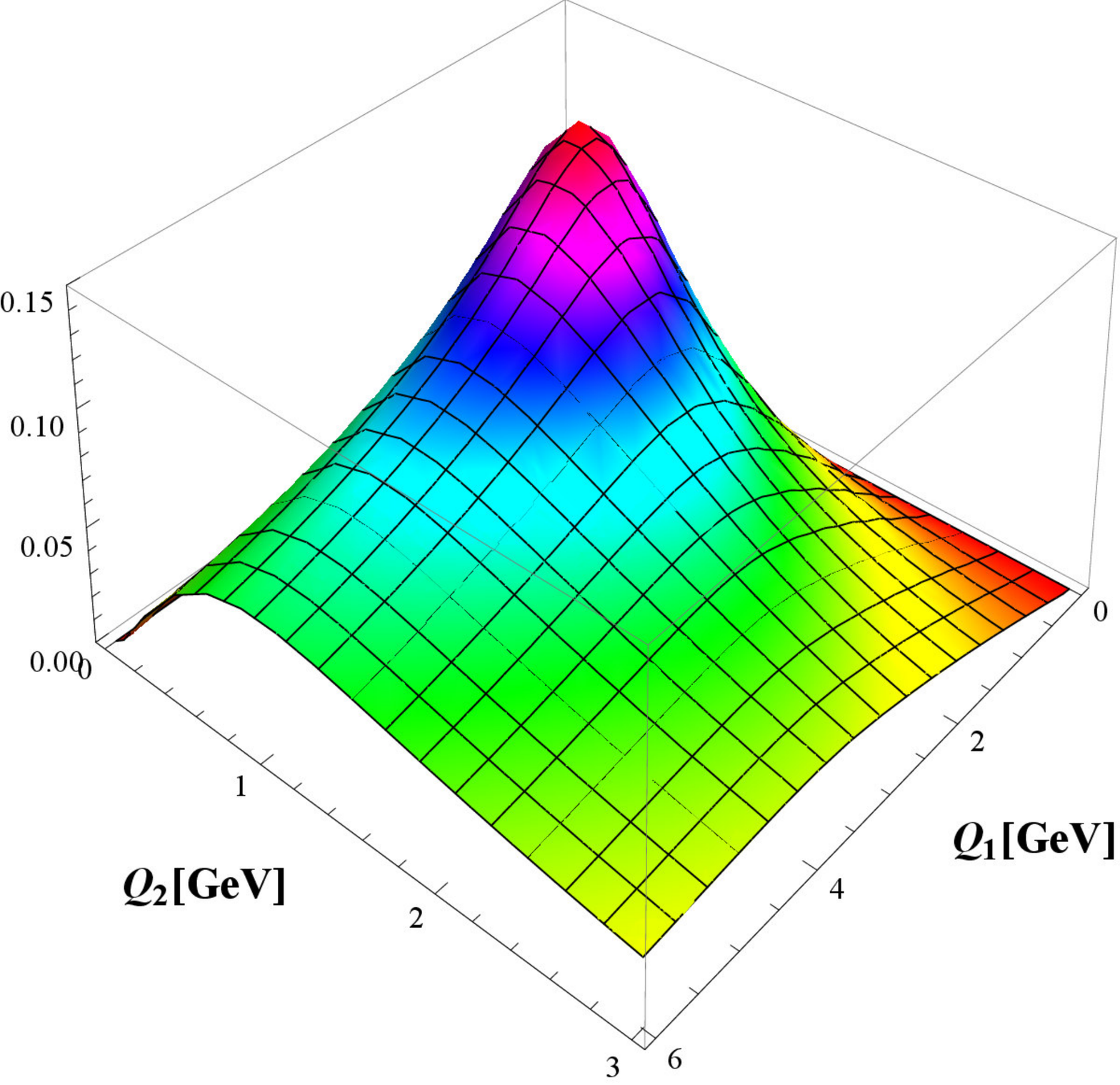}}
{\includegraphics[width=3.5cm  ]{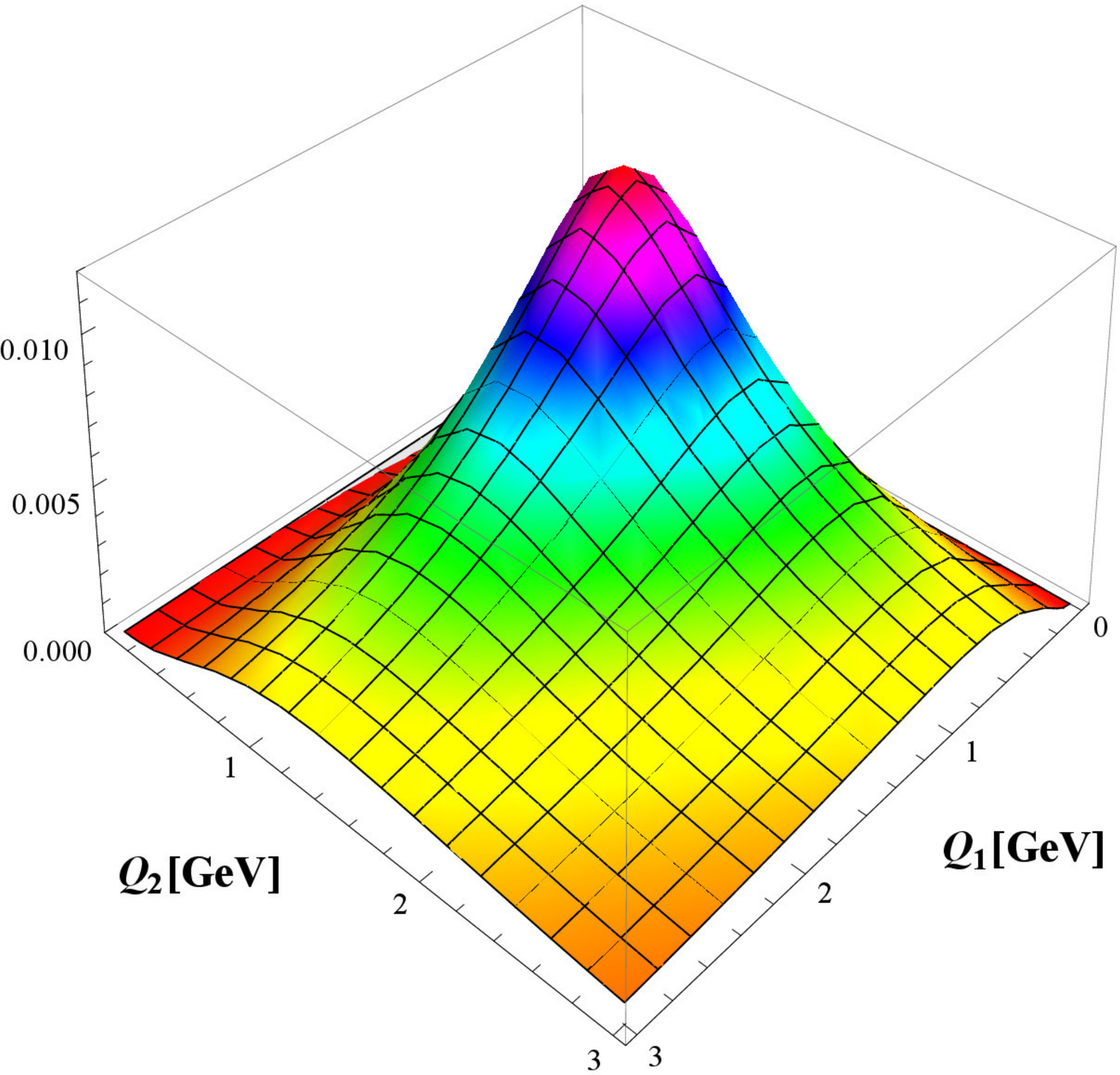}}
{\bf a)}\hspace*{5cm}  {\bf b)}
\caption{The same as in Fig. \ref{fig:weight1} but for the combinations $w_{12}^{PP}$ and $\tilde w_{12}^{PP}$ in Eq. \rf{eq:w12}.}
\label{fig:weight3}
\end{center}
\end{figure}
 %%%%%%%%%%%%%%%%%%%%%%%%%%%%%%%%%%%%%%%
 
\noindent
The dimensionless densities (the overall sign has been chosen such that these
densities are positive) occurring in these expressions can be found in Table \ref{tab:weight1}.
They are obtained upon using the angular integrals given in \cite{KN}. 
Some of their combinations are plotted in Figs. \ref{fig:weight1}, \ref{fig:weight2}, and \ref{fig:weight3}.
Generically, they are peaked in a region around $Q_1 \sim Q_2 \sim 500$ MeV, and are suppressed for
smaller or larger values of the Euclidian loop momenta.

%%%%%%%%%%%%%%%%%%%%%%%%%%%%%%
\section{I=0 Scalar Mesons from Gluonium Sum Rules}
\label{sec:scalars} 
\setcounter{equation}{0}
%%%%%%%%%%%%%%%%%%%%%%%%%%%%%%

The evaluation of $a_\mu^{lbl}\vert_S^{\rm VMD}$ as given in Eq. \rf{a_mu_integrals},
requires as input values for the masses amd the two-photon widths of the various
scalar resonances we want to include. For the narrow states, this information
can be gathered from the review \cite{RPP18} or from other sources, which will
be described in Section \ref{sec:scalar_widths}.
In this section, we review the information provided  by various QCD spectral sum rules and some low-energy theorems
on the mass, as well as on the hadronic and two-photon widths,
of the lightest scalar meson $\sigma/f_0(500)$, the $f_0(1350)$ and $f_0(1504)$ interpreted as
gluonia states.
%%%%%%%%%%%%%%%%%%%%%%%%%%%%%%%%%%%%%%%%%
\subsection*{\b I=0 Scalar Mesons as gluonia candidates}
%%%%%%%%%%%%%%%%%%%%%%%%%%%%%%%%%%%%%%%%%
The nature of the isoscalar $I=0$ scalar states remains unclear as it goes beyond the usual octet quark model description due to their $U(1)$ component. A four-quark description of these states have been proposed within the bag model\,\cite{JAFFE} and studied phenomenologically in e.g Refs. \,\cite{ACHASOV2,THOOFT}. However, its singlet nature has also motivated their interpretation as gluonia candidates as initiated in Ref.\,\cite{MIN} and continued in Refs.\,\cite{MIN2,NSVZ,SNV,SNGLUE,SNPRD}\,\footnote{For recent reviews on the experimental searches and on the theoretical  studies of gluonia, see e.g. Refs.\,\cite{OCHS2,KLEMPT,VENTO,CREDE}.}. Recent analysis of the $\pi\pi$ and $\gamma\gamma$ scattering data indicates an eventual large gluon component of the $\sigma/f_0(500)$ and $f_0(990)$ states\,\cite{MENES,OCHS,WANG,WANG2,KAMINSKI} while recent data analysis from central productions\,\cite{GASTALDI} shows the gluonium nature of the $f_0(1350)$ decaying into $\pi^+\pi^-$ and into the specific $4\pi^0$ states via two virtual $\sigma/f_0(500)$ states as expected if it is a gluonium\,\cite{SNV,SNGLUE}.  The $\sigma/f_0(500)$ are  observed in the gluonia golden $J/\psi$ and $\Upsilon \to\pi\pi\gamma $ radiative decays but often interpreted as S-wave backgrounds due to its large width (see e.g BESIII\,\cite{BESIII} and BABAR\,\cite{BABAR}). The glueball nature of the $G(1.5-1.6)$ has been also found by GAMS few years ago\,\cite{GAMS} on its decay to $\eta'\eta$ and on the value of the branching ratio $\eta'\eta$/ $\eta\eta$ expected for a high-mass gluonium\,\cite{SNV,SNGLUE}.  
%%%%%%%%%%%%%%%%%%%%%%%%%%%
\subsection*{\b The $\sigma/f_0$ mass from QCD spectral sum rules}
%%%%%%%%%%%%%%%%%%%%%%%%%%%
The singlet nature of the $\sigma/f_0$ has motivated to consider that it may contain a large gluon component
 \cite{NSVZ,SNV,SNGLUE}, which may explain its large mass compared to the pion. This property is encoded in the 
 trace of the QCD energy momentum tensor:
\beq
\theta^\mu_\mu=\frac{1}{4}\beta(\alpha_s)G^a_{\mu\nu}G^{\mu\nu}_a +\big{[} 1+\gamma_m(\alpha_s)\big{]}\sum_{u,d,s}m_q\bar\psi_q\psi_q,
\eeq
where $\beta(\alpha_s)\equiv \beta_1(\alpha_s/\pi)+\cdots$ and $\gamma_m(\alpha_s)\equiv \gamma_1(\alpha_s/\pi)+\cdots$ are the 
QCD $\beta$-function and quark mass anomalous dimension: $-\beta_1=(1/2)(11-2n_f/3)$, $\gamma_1=2$ for $SU(3)_c\times SU(n_f)$. 
A QCD spectral sum rule (QSSR)\,\cite{SVZa,SVZb}\,\footnote{For reviews, see the textbooks in Refs.\,\cite{SNB1,SNB2} and 
reviews in Refs.\,\cite{SNB3,SNB4}.} analysis of the corresponding two-point correlator in the chiral limit ($m_q=0$):
\beq
\psi_g(q^2)=i\int d^4x \la 0|{\cal T} \theta^\mu_\mu(x)\theta^\mu_\mu(0)|0\ra
\eeq
from the subtracted and unsubtracted Laplace sum rules:
\bea
{\cal L}_0(\tau)&=&\int_0^\infty dt e^{_t\tau}\frac{1}{\pi}{\rm Im}\psi_g(t)\nnb\\
{\cal L}_{-1}(\tau)&=&-\psi_g(0)+\int_0^\infty\frac {dt}{ t}e^{_t\tau}\frac{1}{\pi}{\rm Im}\psi_g(t)
\eea
leads to the predictions 
\beq
M_{\sigma}\approx (0.95-1.10)~{\rm GeV}~~~{\rm and } ~~~M_{G}\approx (1.5-1.6)~{\rm GeV}
\label{eq:srmass}
\eeq
for the masses of the $\sigma/f_0$ and scalar gluonium states.
%suggesting that the lowest gluonium mass may be indentified with the on-shell mass of the $\sigma$ meson.

%%%%%%%%%%%%%%%%%%%%%%%%%%%%%%%%%%%%%%%%%%
\begin{table*}[hbt]
\setlength{\tabcolsep}{1.pc}
\caption{Expressions of the weight functions defined in Eq. \rf{eq:w12} after angular integration in the Euclidian space
[$D_{m1} \equiv (P + Q_1)^2+m^2$, $D_{m2} \equiv (P - Q_2)^2+m^2$].  }
\begin{tabular*}{\textwidth}{@{}r@{\extracolsep{\fill}}l}
&\\
\hline
&\\
$ {w_1^{PP}}(M)$=& $ - \int \frac{d \Omega_1}{2 \pi^2}  \int \frac{d \Omega_2}{2 \pi^2}
\frac{\pi^2 Q_1 Q_2}{D_{m1} D_{m2}} 
 \frac{T_{1E}^{S;PP} (Q_1 , Q_2)}{(Q_2^2 + M_S^2) [(Q_1 + Q_2)^2 + M^2]}$\\
 =&
$-\frac{2}{3}\frac{\pi ^2 { Q_1}  { Q_2}}{ { Q_2}^2+ {M_S}^2}\Big{[}1+
\frac{ { Q_2}^2}{2 m_l^2}+ { Q_2}^2\left( { Q_1}^2- { Q_2}^2- M^2\right)\left( { Q_1}^2- { Q_2}^2- M^2-4 m_l^2\right) {I_1^M}
-\Big{(}2 { Q_1}^2- { Q_2}^2- M^2+\frac{ { Q_1}^2 { Q_2}^2}{2 m_l^2}\Big{)} \frac{ {Rm_1}-1}{2 m_l^2}$\\
&$- { Q_2}^2\left(2+\frac{ { Q_2}^2}{2 m_l^2}\right)\frac{ {Rm_2}-1}{2 m_l^2} -
\left( { Q_1}^2- { Q_2}^2+ M^2\right) {I_7^M} + { Q_2}^2 \Big{(}3 { Q_1}^2- { Q_2}^2- M^2-4 m_l^2 + \frac{ { Q_1}^2  { Q_2}^2}{2 m_l^2}\Big{)}
\frac{ {Rm_1}-1}{2 m_l^2} \,\frac{ {Rm_2}-1}{2 m_l^2}$\\
&$+\left( { Q_1}^2+ { Q_2}^2+ M^2\right)\left(2 { Q_1}^2- { Q_2}^2- M^2\right)\frac{ {Rm_1}-1}{2 m_l^2} {I_7^M}-2 { Q_2}^2
\left( { Q_1}^2- M^2\right)\frac{ {Rm_2}-1}{2 m_l^2} {I_7^M}\Big{]},$\\
\\ 

${w_1^{PQ}}(M)$=&$- \int \frac{d \Omega_1}{2 \pi^2}  \int \frac{d \Omega_2}{2 \pi^2}
\frac{\pi^2 Q_1 Q_2}{D_{m1} D_{m2}}  \frac{1}{M_S^2} \frac{T_{1E}^{S;PQ} (Q_1 , Q_2)}{(Q_2^2 + M_S^2) [(Q_1 + Q_2)^2 + M^2]} $\\
=&$-\frac{1}{3}\frac{\pi ^2{Q_1} {Q_2}}{{M_S}^2\left({Q_2}^2+{M_S}^2\right)}\Big{[}{Q_1}^2+{Q_2}^2-M^2+2\frac{{Q_1}^2{Q_2}^2}{m_l^2}-4{Q_1}^2{Q_2}^2M^2\left({Q_1}^2-2{Q_2}^2-M^2-4m_l^2\right){I_1^M}-{Q_1}^2\big{(}{Q_1}^2$\\
&$-3{Q_2}^2-5M^2+2\frac{{Q_1}^2{Q_2}^2}{m_l^2}\big{)}\frac{{Rm_1}-1}{2m_l^2}
-4{Q_2}^2\left(2{Q_1}^2+\frac{{Q_1}^2{Q_2}^2}{2m_l^2}\right)\frac{{Rm_2}-1}{2m_l^2}
-\left({Q_1}^2-{Q_2}^2+M^2\right)\left({Q_1}^2-{Q_2}^2-M^2\right){I_7^M}
$\\
&$+4{Q_1}^2{Q_2}^2\left(2{Q_1}^2-{Q_2}^2-M^2-4m_l^2+\frac{{Q_1}^2{Q_2}^2}{2m_l^2}\right)
\frac{{Rm_1}-1}{2m_l^2}\frac{{Rm_2}-1}{2m_l^2}+{Q_1}^2\Big{(}\left({Q_1}^2-{Q_2}^2\right)^2-4M^2{Q_1}^2
-8{Q_2}^2M^2$\\
&$-5M^4\Big{)}\frac{{Rm_1}-1}{2m_l^2}{I_7^M}
+8M^2{Q_1}^2{Q_2}^2\frac{{Rm_2}-1}{2m_l^2}{I_7^M}\Big{]}, $\\
\\

$w_2^{PP}(M)$=& $ + \int \frac{d \Omega_1}{2 \pi^2}  \int \frac{d \Omega_2}{2 \pi^2}
\frac{\pi^2 Q_1 Q_2}{D_{m1} D_{m2}} \frac{T_{2E}^{S;PP} (Q_1 , Q_2)}{[(Q_1 + Q_2)^2 + M_S^2] [(Q_1 + Q_2)^2 + M^2]} 
\equiv\frac{\tilde{w}_2^{PP}(M)-\tilde{w}_2^{PP}(M_S)}{M_S^2-M^2}$,\\
\\

$ w_2^{PQ}(M)$=&$- \int \frac{d \Omega_1}{2 \pi^2}  \int \frac{d \Omega_2}{2 \pi^2}
\frac{\pi^2 Q_1 Q_2}{D_{m1} D_{m2}} \frac{1}{M_S^2}\frac{T_{2E}^{S;PQ} (Q_1 , Q_2)}{[(Q_1 + Q_2)^2 + M_S^2] [(Q_1 + Q_2)^2 + M^2]}\equiv
\frac{\tilde{w}_2^{PQ}(M)-\tilde{w}_2^{PQ}(M_S)}{M_S^2(M_S^2-M^2)}$,\\
\\
with~~~~~~~~ & \\
\\ 
%$D_{m1} \equiv (P + Q_1)^2+m^2,\quad D_{m2} \equiv (P - Q_2)^2+m^2$,\\
${\tilde{w}_2^{PP}}(M)$=&$\frac{2}{3}\pi ^2{Q_1} {Q_2}\Big{(}2M^2\big{(}{Q_1}^2 {Q_2}^2+m_l^2{Q_1}^2+m_l^2{Q_2}^2+m_l^2M^2\big{)}{I_1^M}+\frac{{Q_1}^2}{2}\frac{{Rm_1}-1}{2m_l^2}+\frac{{Q_2}^2}{2}\frac{{Rm_2}-1}{2m_l^2}+M^2{I_7^M}$\\
&$-\left({Q_1}^2{Q_2}^2+2m_l^2M^2\right)\frac{{Rm_1}-1}{2m_l^2}\frac{{Rm_2}-1}{2m_l^2}-\frac{{Q_1}^2}{2}\left({Q_1}^2-{Q_2}^2+3M^2\right)\frac{{Rm_1}-1}{2m_l^2}{I_7^M}-\frac{{Q_2}^2}{2}\left({Q_2}^2-{Q_1}^2+3M^2\right)\frac{{Rm_2}-1}{2m_l^2}{I_7^M}\Big{)}, $\\
\\

${\tilde{w}_2^{PQ}}(M)$=&$-\frac{1}{3}{\pi ^2}{Q_1} {Q_2}\Big{[}-M^2+2M^2{Q_1}^2{Q_2}^2({Q_1}^2+{Q_2}^2+4m_l^2){I_1^M}+\frac{{Q_1}^2}{2}\left({Q_1}^2+3{Q_2}^2+M^2\right)\frac{{Rm_1}-1}{2m_l^2}$\\
&$+\frac{{Q_2}^2}{2}({Q_2}^2+3{Q_1}^2+M^2)\frac{{Rm_2}-1}{2m_l^2}+M^2({Q_1}^2+{Q_2}^2+M^2){I_7^M}-2{Q_1}^2{Q_2}^2\left({Q_1}^2+{Q_2}^2+4m_l^2\right)\frac{{Rm_1}-1}{2m_l^2}\frac{{Rm_2}-1}{2m_l^2}-\frac{{Q_1}^2}{2}$\\
&$\times({Q_1}^4-{Q_2}^4+2M^2{Q_1}^2+4M^2{Q_2}^2+M^4)\frac{{Rm_1}-1}{2m_l^2}{I_7^M}-\frac{{Q_2}^2}{2}({Q_2}^4-{Q_1}^4+2M^2{Q_2}^2+4M^2{Q_1}^2+M^4)\frac{{Rm_2}-1}{2m_l^2}{I_7^M}\Big{]}$,\\
%&$\Delta w_1^{PP,PQ}(M) \equiv w_1^{PP,PQ} (M) - w_1^{PP,PQ} (0)\,,  \quad\quad
% \Delta w_2^{PP,PQ}(M) \equiv w_2^{PP,PQ} (M) - w_2^{PP,PQ} (0)\, $\\
and~~~~~~~~~~&
$ I_1^M=\frac{1}{m_l^2 Q_1^2 Q_2^2}\ln [1-Z_{Q_1Q_2}^MZ_{PQ_1}^{m_l}Z_{PQ_2}^{m_l}],\quad\quad I_7^M= \frac{Z_{Q_1Q_2}^M}{Q_1 Q_2}, \quad\quad R_{mi}\equiv \sqrt{1+\frac{4m_l^2}{Q_i^2}}, \quad\quad Z_{PQ_i}^{m_l}=\frac{Q_i}{2P}(1-R_{mi}),$\\
&$(Z_{PQ_i}^{m_l})^2=\frac{Q_i}{P}Z_{PQ_i}^{m_l}-1,\quad\quad   Z_{PQ_1}^{m_l}Z_{PQ_2}^{m_l}=-\frac{Q_1 Q_2}{4m_l^2}(R_{m1}-1)((R_{m2}-1)), \quad\quad Z_{KL}^M=\frac{K^2+L^2+M^2-\sqrt{(K^2+L^2+M^2)^2-4K^2L^2}}{2KL},$\\
\hline
\end{tabular*}
\label{tab:weight1}
\end{table*}
%%%%%%%%%%%%%%%%%%%%%%%%%%%%%%%%%%%%%%%

%%%%%%%%%%%%%%%%%%%%%%%%%%
\subsection*{ \b $\sigma/f_0$ hadronic width from vertex sum rules}
%%%%%%%%%%%%%%%%%%%%%%%%%%
The $\sigma$ hadronic width can be estimated from the  vertex function:
\beq
V\big{[}q^2\equiv (q_1-q_2)^2\big{]}=\la\pi | \theta^\mu_\mu |\pi\ra,
\eeq
which obeys a once subtracted dispersion relation \cite{SNV,SNGLUE}:
\beq
V(q^2)=V(0)+q^2\int_{4m_\pi^2}^{\infty}\frac{dt}{t}\frac{1}{ t-q^2-i\epsilon}\frac{1}{ \pi}{\rm Im} V(t)~
\eeq
From the  low-energy constraints:
\beq
V(0)={\cal O}(m_\pi^2)\to 0,~~~~~V'(0) = 1,
\eeq
one can derive the low-energy sum rules :
\beq
\frac{1}{ 4}\sum_{S\equiv \sigma,\dots}g_{S\pi\pi}\sqrt{2}f_S=0,~~~\frac{1}{ 4}\sum_{S\equiv \sigma,\dots}g_{S\pi\pi}\frac{\sqrt{2}f_S}{ M_S^2}=1,
\eeq
where $f_S$ is the scalar decay constant normalized as %$f_\pi=93$ MeV:
\beq
\la 0\vert 4\theta^\mu_\mu\vert S\ra=\sqrt {2} f_S M_S^2~,
\eeq
with \cite{SNGLUE}:
\beq
f_{\sigma}\simeq 1~{\rm GeV},~~~f_{\sigma'}\simeq 0.6~{\rm GeV},~~~f_{G}\simeq 0.4~{\rm GeV},
\eeq
 for $M_{\sigma}\simeq 1$ GeV, $M_{\sigma'}\simeq 1.3$ GeV and $M_G\simeq 1.5$ GeV.
 The first sum rule requires the existence of two resonances, $\sigma/f_0$ and its radial exictation $\sigma'$, 
 coupled strongly to $\pi\pi$\,\footnote{The $G(1600)$ is found to couple weakly to $\pi\pi$ and might
 be identified with the gluonium state obtained in the lattice quenched approximation (for a recent review of different lattice results, see e.g.\,\cite{OCHS2}). }. 
 Solving the second sum rule gives, in the chiral limit,
\beq
|g_{\sigma\pi^+\pi^-}|\simeq |g_{\sigma K^+K^-}|\simeq (4-5)~{\rm GeV},
\eeq
which suggests an  universal coupling of the $\sigma/f_0$ to Goldstone boson pairs as confirmed from 
the $\pi\pi$ and $\bar KK$ scatterings data analysis\,\cite{WANG2,KAMINSKI}. This result leads to the hadronic width:
\beq
 \Gamma_{\sigma\to\pi\pi}\equiv \frac{|g_{\sigma\pi^+\pi^-}|^2}{16\pi M_{\sigma}}\ga 1-\frac{4m_\pi^2}{ M^2_{\sigma}}\dr^{1/2}\approx 0.7 ~{\rm GeV}.
 \eeq 
 This large width into $\pi\pi$ is a typical OZI-violation expected to be due to large non-perturbative 
 effects in the region below 1 GeV. Its value compares quite well with the width of the so-called on-shell 
 $\sigma/f_0$ mass obtained in Ref.\,\cite{OCHS,WANG,WANG2} (see also the next subsection).
  %%%%%%%%%%%%%%%%%%%%%%%%%%%%%%%%%%%%%
\subsection*{ \b $\sigma/f_0\to \gamma\gamma$ width from some low-energy theorems}
%%%%%%%%%%%%%%%%%%%%%%%%%%%%%%%%%%%%%5
 We introduce the gauge invariant scalar meson coupling to $\gamma\gamma$ through the interaction Lagrangian and  related coupling:
\be
{\cal L}_{int}=
\frac{g_{S\gamma\gamma}}{2} F_{\mu\nu} F^{\mu\nu}~,~~~~ {\cal P}(0,0)\equiv \tilde g_{S\gamma\gamma}=\ga \frac{2}{e^2}\dr g_{S\gamma\gamma}~,
\lbl{eq:int}
\ee
where $F_{\mu\nu}$ is the photon field strength. In momentum space, the corresponding interaction 
reads\,\footnote{We use the normalization and structure in\,\cite{RENARD} for on-shell photons. However, a more general expression is presented in\,\cite{MS} for off-shell photons. We plan to come back to this point in a future publication.
}
\beq
{\cal L}_{int}=2g_{S\gamma\gamma} P_{\mu\nu} (q_1 q_2) \times \epsilon_1^\mu \epsilon_2^\nu,
\lbl{eq:vertex}
\eeq
where $\epsilon_i^\mu$ are the photon polarizations. With this normalization, the decay width reads
\beq
\Gamma = \vert g_{S\gamma\gamma}\vert^2\frac{M_S^3}{8\pi} \ga \frac{1}{2} \dr
= \frac{\pi}{4} \alpha^2 M_S^3 \vert {\tilde g}_{S\gamma\gamma}\vert^2,
\eeq
where 1/2 is the statistical factor for the two-photon state. 
One can for instance estimate the $\sigma \gamma\gamma$ coupling by identifying the Euler-Heisenberg
Lagrangian derived from $gg\to  \gamma\gamma$ via a quark constituent loop with the interaction Lagrangian in Eq. \rf{eq:int}.
%\beq
%{\cal L}_{S\gamma\gamma}=g_{S\gamma\gamma}SF_{\mu\nu}^{(1)}F_{\mu\nu}^{(2)}.
%\eeq
In this way, one deduces the constraint\,\footnote{This sum rule has been originally used by \cite{NSVZ} in the case 
of a charm quark loop for estimating the $J/\psi\to\gamma\sigma$ radiative decay.}:
\beq
g_{S\gamma\gamma}\simeq \frac{\alpha}{ 60}\sqrt{2}f_SM^2_S\ga \frac{\pi}{-\beta_1}\dr\sum_{u,d,s} Q_q^2/M^4_q,
\eeq
%\end{document}
where $Q_q$ is the quark charge in units of $e$; $M_{u,d}\approx M_\rho/2$ and $M_\phi\approx M_\phi/2$ are constituent quark masses. 
Then, one obtains:
\beq
g_{\sigma\gamma\gamma}\approx g_{\sigma'\gamma\gamma}\approx g_{G\gamma\gamma}\approx (0.4-0.7)\alpha~{\rm GeV}^{-1},
\lbl{eq:gam-gam}
\eeq
which leads, for $M_\sigma \simeq 1$ GeV, to the $\gamma\gamma$ width:
\beq
\Gamma_{\sigma\to  \gamma\gamma}\simeq (0.2-0.6)~{\rm keV}.
\label{eq:sr2gamma}
\eeq
A consistency check of the previous result can be obtained from the trace anomaly $\la 0|\theta_\mu^\mu|\gamma\gamma\ra$ 
by matching the $k^2$ dependence of its two sides which leads to\,\cite{CREW,CHANO,DIVECCHI,LANIK} :
\beq
\frac{1}{ 4}\sum_{S=\sigma\cdots}g_{S\gamma\gamma}\sqrt{2}f_S=\frac{\alpha R}{ 3\pi},
\eeq
where $R\equiv 3\sum Q_q^2$.

%%%%%%%%%%%%%%%%%%%%%%%%%%%%%%
\section{ $\sigma/f_0(500)$ meson from $\pi\pi$ and $\gamma\gamma$ scattering}
\label{sec:sigma}
\setcounter{equation}{0}
%%%%%%%%%%%%%%%%%%%%%%%%

The mass and the width of a broad resonance like the $\sigma/f_0(500)$ state
in general turn out to be rather ambiguous quantities. A non ambiguous definition
is provided by the location of the pole of the S-matrix amplitude on the
second Riemann sheet \cite{resonances}. The difficulty then lies in relating
this pole in the complex domain to the description, for instance in the form of a 
Breit-Wigner function, of the data on the positive real axis. This issue 
has been quite extensively discussed in the context of the line-shapes
of the electroweak gauge and scalar bosons \footnote{The issue was mainly centered
around the necessity to define gauge-invariant observables and to correctly account
for threshold effects. %The first aspect is clearly not a concern in our case.
}
\cite{VALENCIA,Sirlin:1991fd,STUART,VELT,SIRLIN2,SIRLIN,Grassi:2000dz}.

In this section, the information on the $f_0/\sigma$ resonance
that can be obtained from data on $\pi\pi$ scattering or on $\gamma\gamma
\to \pi^0\pi^0,~\pi^+\pi^-$ are reviewed. We then end this section by specifying how the
contribution to HLbL from a broad object like the $\sigma/f_0(500)$ can be
described by the formalism that we have set up in Section \ref{sec:scalar_lbl}.

\subsection*{\b $\sigma/f_0$ mass  and  width in the complex plane}
%%%%%%%%%%%%%%%%%%%%%%%%
The mass and width of the $\sigma/f_0$ meson play an important r\^ole in the present analysis. 
Their precise determinations in the complex plane from $\gamma\gamma
\to \pi^0\pi^0,~\pi^+\pi^-$ scattering data in Ref. \cite{OCHS} (one resonance $\oplus$ one channel) 
and in Refs. \cite{WANG,WANG2} (two resonances $\oplus$ two channels and adding the $K_{e4}$ data), lead to the complex pole:
\bea
M_\sigma^c &\equiv& M_\sigma -i\Gamma_\sigma/2, \nnb\\
&\simeq& \big{[}452(12)-{\rm i}260(15)\big{]}~{\rm MeV},
\lbl{eq:sigma}
\eea
which agrees with some other estimates from $\pi\pi$ scattering data for one channel\,\cite{LEUTa,LEUTb,YND}. Using the model 
of \cite{MENES} for separating the direct and rescattering contributions, one obtains from $\gamma\gamma\to\pi\pi$ 
scatterings data \cite{OCHS,WANG,WANG2}:
\bea
\Gamma_{\sigma}^{\gamma\gamma}\vert^{dir}&\simeq& (0.16\pm 0.04)~{\rm keV},\nnb\\\Gamma_{\sigma}^{\gamma\gamma}\vert^{resc}&\simeq& (1.89\pm 0.81)~{\rm keV},\nnb\\
\Gamma_{\sigma}^{\gamma\gamma}\vert^{tot}&\simeq &(3.08\pm 0.82)~{\rm keV},
%\lbl{eq:gamma}
\label{eq:gamma}
\eea
corresponding respectively to the {\it direct, rescattering} contributions and {\it their total sum}. The rescattering contribution 
includes the ones of the Born term, the vector and axial-vector mesons in the $t$-channel and the $I=2$ mesons. 
%%%%%%%%%%%%%%%%%%%%%%%%
\subsection*{\b $\sigma/f_0$ Breit-Wigner on-shell mass  and  widths}
%%%%%%%%%%%%%%%%%%%%%%%%
However, an extrapolation of the previous result obtained in the complex plane to the real axis is not straightforward. 
Then, in the Breit-Wigner analysis for approximately reproducing the data, one may either introduce the {\it on-shell mass and width} defined in
\cite{SIRLIN} for the $Z$-bozon and used  \cite{OCHS,WANG2,OCHS2} within the model of \cite{MENES}:
\beq
{\rm Re}~{\cal D}[(M_\sigma^{\rm os})^2]=0~\lrar~
M_\sigma^{\rm os}\approx 0.92 ~{\rm GeV} ~.
\eeq
It corresponds to the {\it on-shell hadronic width} evaluated at $s=(M_\sigma^{\rm os})^2$:
\beq
M_\sigma^{\rm os}\Gamma^{\pi\pi}_{\sigma}\vert_{os}\simeq\frac{{\rm Im}~{\cal D}}{-{\rm Re}~{\cal D}'}
~\lrar 
\Gamma^{\pi\pi}_{\sigma}\vert_{os}\approx 1.04~{\rm GeV}, 
%\lbl{eq:mass-onshell}
\label{eq:mass-onshell}
\eeq
where ${\cal D}$ is the inverse propagator and ${\cal D}'$ its derivative. The corresponding  $\gamma\gamma$ width can be extracted by evaluating  
Eq. (\ref{eq:gamma}) at the on-shell mass and gives by including the $f_0(980)$ in the fit analysis\,\cite{WANG2}:
\beq
\Gamma_{\sigma}^{\gamma\gamma}\vert_{os}\simeq (1.2\pm 0.3)~{\rm keV}.
%\lbl{eq:gam-onshell}
\label{eq:gam-onshell}
\eeq
A more recent fit of the data using the Breit-Wigner parametrization leads to\,\cite{OCHS2}:
\beq
M_{\sigma}\simeq 1000(100)~{\rm MeV},~~~~~~~~\Gamma^{\pi\pi}_{\sigma}\simeq 700(70)~{\rm MeV}~,
\eeq
which are consistent with the above results, and with the sum rules results in Eq.\,(\ref{eq:srmass}). An earlier fit using K-matrix leads to the value\,\cite{PEN} :  
\beq
M_\sigma= 910-350~{\rm i} ~{\rm MeV}~,
\eeq
quoted without errors. 
%%%%%%%%%%%%%%%%%%%%%%%%%%%%%%%%%%%%%
\subsection*{\b %Continuation of a 
Breit-Wigner function in the space-like domain}
%%%%%%%%%%%%%%%%%%%%%%%%%%%%%%%%%%%%%

\begin{figure}[ht]
\begin{center}
{\includegraphics[width=7cm]{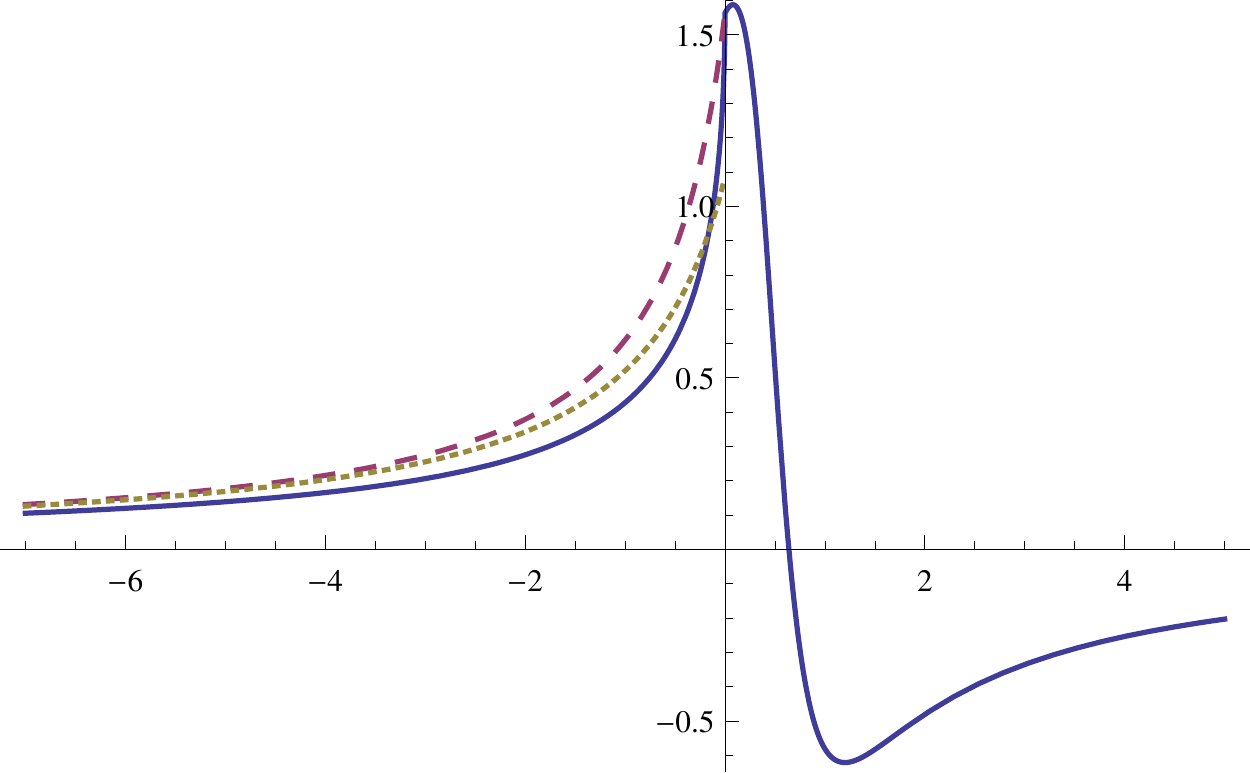}}\\
\caption{The function ${\widetilde{BW}} (s;M_{\rm BW},\Gamma_{\rm BW})$ 
(solid line) for 
$M_{\rm BW}=0.8$ GeV and $\Gamma_{\rm BW}=0.7$ GeV, as a function of $s$
(in GeV$^2$), compared, for negative values of $s$, to the function $-1/(s-M_{\rm BW}^2)$
(dashed line), for the same value of $M_{\rm BW}$, and to the function
$-1/(s-M_{\rm eff}^2)$ (dotted line), with $M_{\rm eff} = 1.2 M_{\rm BW}$,
which gives a better description in the region around $s\sim (0.5~{\rm GeV})^2$.}
\label{fig:BW}
\end{center}
\end{figure}

Let us assume that the data on the real positive axis are described in terms
of a Breit-Wigner function $BW(s;M_{\rm BW},\Gamma_{\rm BW})$ for some
values of the Breit-Wigner mass $M_{\rm BW}$ and width $\Gamma_{\rm BW}$.
In order to extend this function on the whole real $s$-axis without introducing
any singularity besides the cut along the positive real axis, one considers
the function \cite{Lomon:2012pn,Moussallam:2013una}:
\be
{\widetilde{BW}} (s;M_{\rm BW},\Gamma_{\rm BW})
= \frac{1}{\pi} \int_0^\infty dx \, \frac{{\rm Im}\,BW(s;M_{\rm BW},\Gamma_{\rm BW})}{x - s - i \epsilon}
.~~
\ee
For:
\be
BW(s;M_{\rm BW},\Gamma_{\rm BW})
=
\frac{1}{s - M_{\rm BW}^2 - i \sqrt{s}\, \Gamma_{\rm BW}},
\ee
one finds ${\widetilde{BW}} (s;M_{\rm BW},\Gamma_{\rm BW}) = BW (s;M_{\rm BW},\Gamma_{\rm BW})$
for $s>0$, and for $s=-Q^2<0$:
\be
{\widetilde{BW}} (-Q^2;M_{\rm BW},\Gamma_{\rm BW}) = 
\frac{-1}{Q^2 + M_{\rm BW}^2 + \sqrt{Q^2} \,\Gamma_{\rm BW}}
.
\ee
In the narrow-width approximation, this reduces to the usual
Euclidian version of the Feynman propagator. But the latter
represents a good approximation even when the width becomes sizeable.
This is illustrated in Fig. \ref{fig:BW} for the case $\Gamma_{\rm BW} \sim M_{\rm BW}$. 
One can also represent the
function ${\widetilde{BW}} (s;M_{\rm BW},\Gamma_{\rm BW})$
in the space-like region by a propagator term $-1/(s-M_{\rm eff}^2)$,
with $M_{\rm eff}$ adjusted, for instance, to give a more
accurate description of ${\widetilde{BW}} (s;M_{\rm BW},\Gamma_{\rm BW})$
in the region of values of $Q^2$ that matters most from the point of
view of the weight functions displayed in Figs.\,\ref{fig:weight1}
and \ref{fig:weight3}. Given the large uncertainties in the mass
of the $\sigma/f_0(500)$, such refinements will actually not be necessary.

%%%%%%%%%%%%%%%%%%%%%%%%%
 %%%%%%%%%%%%%%%%%%%%%%%%%%%%%%%%%%%%%%%%%%%%%%%%%%%%%%%%%%%%%%%%

%%%%%%%%%%%%%%%%%%%%%%%%%
%\begin{figure}[ht]
%\begin{center}
%\vspace{-7cm}
%~~~~{\includegraphics[width=8cm]{plot_BWL.pdf}}\\
%\vspace{-8cm}
%~~~~{\includegraphics[width=8cm]{plot_BWR.pdf}}\\
%\caption{ Light-by-light Hadron scattering contribution to $a_l$. The wavy lines represent photon. The cross correponds to the insertion of the electromagnetic current. 
%The shaded box represents hadrons subgraphs.}
%\label{fig:box}
%\end{center}
%\end{figure}
 %%%%%%%%%%%%%%%%%%%%%%%%%%%%%%%%%%%%%%%%%%%%%%%%%%%%%%%%%%%%%%%%

 %%%%%%%%%%%%%%%%%%%%%%%%%%%%%%%%%%%%%
\section{Adopted values of the $\sigma/f_0(500)$ mass and widths}
\label{sec:sigma_values}
\setcounter{equation}{0}
%%%%%%%%%%%%%%%%%%%%%%%%%%%%%%%%%%%%%
\subsection*{\b $\sigma/f_0(500)$ mass and hadronic width}
%%%%%%%%%%%%%%%%%%%%%%%%%%%%%%%%%%%%%%
Assuming that the realtive errors in the fitting procedure of Ref.\,\cite{PEN} are the same as the ones in Ref.\,\cite{OCHS2} and taking the range of values spanned by the three different determinations including the sum rules
results inEq.\,\ref{eq:srmass}, we adopt the values:
\beq
M_{\sigma}\simeq (960\pm 96),~{\rm MeV}~~~~~~~~\Gamma^{\pi\pi}_{\sigma}\simeq(700\pm 70)~{\rm MeV},
\label{eq:msigma}
\eeq
which implicitly  includes in its definition the large hadronic width of the $\sigma$-meson. One should notice that the three predictions for the widths agree each other and we have assumed a guessed error of 10\%. 

%It is not straightforward to choose among the two above proposals for the real axis. Our choice will be guided by comparing these results with the ones from the PDG values. 
We compare the previous values with the range given by PDG\,\cite{RPP18} for a Breit-Wigner (BW) mass and hadronic width (in units of MeV):
\beq
M_{\sigma}\simeq (400- 550)~,~~~~~~~~\Gamma^{\pi\pi}_{\sigma}\simeq (400- 700)~,
%~~\Gamma_{\sigma}^{\gamma\gamma}\simeq (1-4)~{\rm keV}. 
\eeq
where we notice that our predictions for the BW mass are slightly higher. 
%%%%%%%%%%%%%%%%%%%%%%%%%%%%%%%%%%%%%%%%%%%%%%%%%%%%%%%%%%%%%%%%%%%%%%%%
 \subsection*{\b $\sigma/f_0(500)\to \gamma\gamma$ width}
%%%%%%%%%%%%%%%%%%%%%%%%%%%%%%%%%%%%%%%%%%%%%%%%%%%%%%%%%%%%%%%%%%%%%%%%%
For the $\gamma\gamma$ width, PDG does not provide any estimated range of values. Among the different estimates proposed in the literature which often refer to the {\it total $\gamma\gamma$-width} of the $\sigma$ in the complex plane, we consider the most recent determinations in Eq.\,(\ref{eq:gamma}) from \,\cite{WANG2} and the ones in 
Refs.\,\cite{HOFER,MOUSS}. Averaging these results with the one in Eq.\,(\ref{eq:gam-onshell}) from \cite{WANG2}, we obtain:
\beq
\Gamma_{\sigma}^{\gamma\gamma}\vert^{tot}_{mean}\simeq (1.82\pm 0.32)~{\rm keV}
\eeq
where we have doubled the error for a conservative result. This  {\it total $\gamma\gamma$-width} is larger than expected from a pure glueball
state\,\cite{SNV,SNGLUE}  indicating the complex dyamics for  extracting the width from the data. The corresponding coupling is:
\beq
\tilde g_{\sigma\gamma\gamma}\equiv\ga\frac{2}{ e^2}\dr g_{\sigma\gamma\gamma}\simeq (0.24\pm 0.02) ~{\rm GeV}^{-1}.
\label{eq:sigma-gamma}
\eeq
%%%%%%%%%%%%%%%%%%%%%%%%%%%%%%%%%%%%%%%%%%%%%%%%%%%%%%%%%%%%%%%%%%%%%%%%

\begin{table*}[hbt]
\begin{center}
%\begin{table*}[hbt]
\setlength{\tabcolsep}{0.18pc}
%\newlength{\digitwidth} \settowidth{\digitwidth}{\rm 0}
%\catcode`?=\active \def?{\kern\digitwidth}

  \caption{\footnotesize    
Scalar mesons contributions to $ a_\mu ^{lbl}\vert_S$ versus their masses. The parameter $\kappa_S$ is defined in Eq.\,\ref{kappa_def}. The errors in the sum have been added quadratically.  
The ${\cal I}_s\,$ integrals with $s\equiv p,pq,q$ are multiplied by $10^2$. 
We use $M_\sigma=(960\pm 96)$ MeV (see text) and $M_V\equiv M_\rho=775$ MeV. 
For the other scalars, the masses are given (in MeV) between parentheses. The errrors on ${\cal I}_{p,...}$ are due to the meson masses. The errors have been added quadratically.}
    {\footnotesize
    \begin{tabular*}{\textwidth}{@{}l@{\extracolsep{\fill}}lllllll}
%\begin{tabular}{llllll}
%   &&&&& \multicolumn{2}{l}{Pilot plant} 
\hline\hline
\\
Scalar&$\tilde g_{S\gamma\gamma}$[GeV$^{-1}$] &$-\mathcal{I}_{p}$[GeV$^{2}$]&$\mathcal{I}_{pq}$[GeV$^{2}$]&$\mathcal{I}_{q}$[GeV$^{2}$]& %
\multicolumn{2}{c}{ $a_\mu ^{lbl}\vert_S\times 10^{11}$}\\
%\cline{6-7} \cline{8-9} 
% &&&&& \multicolumn{1}{c}{ ${\cal Q}(0,0)=0$} 
%                 & \multicolumn{1}{c}{$M_S^2{\cal Q}(0,0)=-\tilde g_{S\gamma\gamma}$} 
  %               & \multicolumn{1}{c}{$M_S^2{\cal Q}(0,0)=2\tilde g_{S\gamma\gamma}$} 
%\\ 
 &&&&& \multicolumn{1}{c}{ $\kappa_S=0$} 
                 & \multicolumn{1}{c}{$\kappa_S=+1$} 
               %  & \multicolumn{1}{c}{$\kappa_S=-2$} 
\\
\hline 
\\
$f_0/\sigma$(960)&$(0.24\pm 0.02)$&$\ga 4.35^{-0.66}_{+0.84}\dr$&$\ga 1.17^{-0.27}_{+0.39}\dr$&$\ga 2.75^{-0.96}_{+1.63}\dr$  %
&$-\ga3.14 ^{-0.72 }_{+0.84}\dr$&$-\ga 0.31 ^{ +0.41}_{-0.82}\dr$\\ %&$+\ga3.11^{-1.99}_{+3.55}\dr$\\
$a_0$(980)&$\ga 0.09 \pm 0.02 \dr$&4.20&1.11&2.51&$-(0.43\pm 0.14)$&$-(0.06 \pm 0.03)$\\ %&$+(0.37\pm {0.16})$\\
$f_0$(990)&$(0.09\pm 0.01)$&4.12&1.08&2.40&$-(0.42\pm 0.09)$&$-(0.07 \pm {0.0 2})$\\%&$+(0.34 \pm 0.08)$\\
$f_{0}$(1350)&$(0.09\pm 0.02)$&2.38&0.44&0.59&$-(0.24\pm 0.11)$&$-(0.14 \pm 0.06)$\\%&${-(0.09\pm 0.04)}$\\
$a_{0}$(1474)&$\ga 0.19^{+0.21}_{-0.08}\dr$&2.03&0.34&0.39&$-\ga 0.92^{+3.15}_{-0.61}\dr$&$-\ga 0.59^{+2.02}_{-0.39}\dr$\\%&$-\ga 0.52^{+1.79}_{-0.34}\dr$\\
$f_{0}$(1504)&$(0.09\pm 0.02)$&1.96&0.32&0.36&$-(0.20\pm 0.09)$&${-(0.13 \pm 0.06)}$\\%&${-(0.12\pm 0.05)}$\\
\\
\hline
\\
{\it Total} &&&&&$-\ga 5.35^{+3.27}_{-0.92}\dr$&$
-\ga 1.3^{+2.06}_{-0.91}\dr$\\%& $+\ga 3.09_{+3.98}^{-2.03}\dr$\\
\\
\hline\hline
\\
%%%%%%%%%%%%%%%%%%%%%%
%\\
%\hline\hline
%\\
%Scalar&$\tilde g_{S\gamma\gamma}$[GeV$^{-1}$] &$-\mathcal{I}_{p}$[GeV$^{2}$]&$\mathcal{I}_{pq}$[GeV$^{2}$]&$\mathcal{I}_{q}$[GeV$^{2}$]& 
%\multicolumn{3}{c}{ $a_\mu ^{lbl}\vert_S\times 10^{11}$} 
%\vspace{0.15cm}
%\\
  %\cline{6-7} \cline{8-9}
% &&&&& \multicolumn{1}{c}{ ${\cal Q}(0,0)=0$} 
%                 & \multicolumn{1}{c}{$M_S^2{\cal Q}(0,0)=-\tilde g_{S\gamma\gamma}$} 
%                 & \multicolumn{1}{c}{$M_S^2{\cal Q}(0,0)=2\tilde g_{S\gamma\gamma}$} 
%\\ 
% &&&&& \multicolumn{1}{c}{ ($\kappa=0$)} 
%                 & \multicolumn{1}{c}{($\kappa=+1$)} 
%                 & \multicolumn{1}{c}{($\kappa=-2$)} 
%\\
%\hline 
%\\
%$f_0/\sigma$(500)&$(0.28\pm 0.04)$&$( 5.85^{-1.18}_{+1.66})$&$( 1.88^{-0.57}_{+0.93})$&$( 6.09^{-2.76}_{+5.82} )$  %
%&$-(5.75 ^{-1.16 }_{+1.63})$&$+(2.09 ^{\bf -2.12}_{+4.99})$&$+(14.51^{-8.61}_{+19.41})$\\
%$a_0$(980)&$\ga 0.09 \pm 0.02 \dr$&4.20&1.11&2.50&$-(0.43 \pm 0.20)$&$-(0.06 \pm 0.03)$&$+(0.37 \pm 0.18)$\\
%$f_0$(990)&$(0.09\pm 0.01)$&4.12&1.07&2.39&$-(0.42\pm 0.10)$&$-(0.07 \pm 0.02)$&$+(0.33 \pm 0.08)$\\
%$f_{0}$(1350)&$(0.19\pm 0.02)$&2.38&0.44&0.59&$-(1.07\pm 0.23)$&$-(0.61 \pm 0.14)$&${-(0.41\pm 0.09)}$\\
%$a_{0}$(1474)&$(0.19^{+0.21}_{-0.08})$&2.03&0.34&0.39&$-(0.92^{+3.15}_{-0.61})$&$-(0.59^{+2.02}_{-0.39})$&$-(0.52^{+1.78}_{-0.35})$\\
%$f_{0}$(1505)&$(0.19\pm 0.02)$&1.95&0.32&0.36&$-(0.88\pm 0.19)$&$-(0.58 \pm 0.13)$&$-(0.53\pm 0.12)$\\\\
%\hline
%\\
%
%%%%%%%%%%%%%%%%%%%%%%
\end{tabular*}
}
\label{tab:amupp}
\end{center}
\end{table*}
 %%%%%%%%%%%%%%%%%%%%%%%%%%%%%%%%%%%%%%%

 %%%%%%%%%%%%%%%%%%%%%%%%%%%%%%%%%%%%%
\section{$\gamma\gamma$  widths of other scalar mesons}
\label{sec:scalar_widths}
\setcounter{equation}{0}
%%%%%%%%%%%%%%%%%%%%%%%%%%%%%%%%%%%%%5%%%%%

%%%%%%%%%%%%%%%%%%%%%%%%%%%%%%%%%%%%%5%%%%%
\subsection*{\b $ f_0(1370)$ and $G\equiv f_0(1500)$ scalar mesons}
%%%%%%%%%%%%%%%%%%%%%%%%%%%%%%%%%%%%%%%%%%%%
 Considering the $ f_0(1370)$ and $G\equiv f_0(1500)$ as gluonium-like scalar mesons \cite{SNV,SNGLUE}, their $\gamma\gamma$ 
 couplings are expected to be given by the sum rule in Eq. \rf{eq:gam-gam}. Then, we take approximately these values to be:
\beq
\tilde g_{\sigma'\gamma\gamma}\approx \tilde g_{G\gamma\gamma}
%\approx \tilde g^{dir}_{\sigma\gamma\gamma}
\simeq (0.09\pm 0.02) ~{\rm GeV}^{-1}.
\eeq
%{\bf
%As I understand it, this should correspond to (6.17) multiplied by
%$2/e^2 = 1/(2\pi\alpha)$, in which case I find
%$$
%\tilde g_{\sigma'\gamma\gamma}\approx \tilde g_{G\gamma\gamma}\approx \tilde g^{dir}_{\sigma\gamma\gamma}
%\simeq (0.06 - 0.11) ~{\rm GeV}^{-1}.
%$$
%}
%%%%%%%%%%%%%%%%%%%%%%%%%
\subsection*{\b $f_0(990)$ scalar meson}
%%%%%%%%%%%%%%%%%%%%%%%%%
The true nature of the $f_0(990)$  is still unclear. However, the large ratio of its coupling $|g_{fK^+K^-}/g_{f\pi^+\pi^-}|\simeq (1.7-2.6) $ 
from $\pi\pi$, $\bar KK$ scatterings  and $J/\psi$-decay data \cite{WANG2,KAMINSKI}  does not favour its $\bar qq$ interpretation but instead 
indicates some gluon or/and four-quark  components. A fit of the $\gamma\gamma$ scattering data leads to the {\it direct width} \cite{WANG2}:
\beq
\Gamma_{f_0}^{\gamma\gamma}\vert^{dir}\simeq 0.28(1)~{\rm keV},
\eeq
which has the same value as the one quoted by PDG \cite{RPP18}:
\beq
\Gamma_{f_0}^{\gamma\gamma}\vert_{PDG}=(0.29\pm 0.07)~{\rm keV},
\lbl{eq:f0dir}
\eeq
from which we deduce the coupling from the {\it direct width}:
\beq
\tilde g_{f_0\gamma\gamma}\simeq (0.09\pm 0.02)~{\rm GeV}^{-1}.
\lbl{eq:gf0dir}
\eeq
One can notice that the {\it rescattering} contribution is large and acts with a destructive interference \cite{WANG2},
\beq
\Gamma_{f_0}^{\gamma\gamma}\vert^{resc}\simeq (0.85\pm 0.05)~{\rm keV}.
\eeq
The ``sum" of the rescattering and direct contributions leads to the {\it  $\gamma\gamma$ total width}
\beq
\Gamma_{f_0}^{\gamma\gamma}\vert^{tot}\simeq (0.16\pm 0.01)~{\rm keV},
\eeq
which is smaller than the direct contribution in Eq. \rf{eq:f0dir}. One can consider 
that the value of the $f_0\to\gamma\gamma$ width is conservatively 
given by the range spanned by the direct and total widths 
\bea
\Gamma_{f_0}^{\gamma\gamma}&=&(0.22\pm 0.07)~{\rm keV}~~~\lrar\nnb\\
&&\tilde g_{f_0\gamma\gamma}\simeq (0.07\pm 0.02)~{\rm GeV}^{-1},
\eea
which is close to the one given in Eq. \rf{eq:f0dir} by PDG. Then, in our analysis, we shall use the PDG value,
which gives:
\beq
\tilde g_{f_0\gamma\gamma}\simeq (0.09\pm 0.01)~{\rm GeV}^{-1}.
\eeq

%%%%%%%%%%%%%%%%%%%%%%%%
\subsection*{\b $a_0(980)$ scalar meson}
%%%%%%%%%%%%%%%%%%%%%%%%
We shall use the value quoted by PDG\,\cite{RPP18}:
\beq
\Gamma_{a_0}^{\gamma\gamma}\ga\frac{\Gamma_{a_0}^{\eta\pi}}
{\Gamma_{a_0}^{tot}}\dr=\ga 0.21^{+ 0.07}_{-0.04}\dr~{\rm keV},
\eeq
where again the rescattering contribution is important \,\cite{ACHASOV}. We deduce:
\beq
\tilde g_{a_0\gamma\gamma}\simeq \ga 0.09^{+0.02}_{- 0.01}\dr~{\rm GeV}^{-1},
\eeq
where we have used : ${\Gamma_{a_0}^{\eta\pi}}/
{\Gamma_{a_0}^{tot}}\simeq 0.82$~\cite{RPP18}.
%%%%%%%%%%%%%%%%%%%%%%%%
\subsection*{\b $a_0(1450)$ scalar meson}
%%%%%%%%%%%%%%%%%%%%%%%%
The origin of the $\gamma\gamma$ width from Belle data on $\gamma\gamma\to \pi^0\eta$ as quoted by the PDG\,\cite{RPP18} is
quite uncertain. Its value is : 
\beq
\Gamma_{a_0}^{\gamma\gamma}\ga\frac{\Gamma_{a_0}^{\eta\pi}}
{\Gamma_{a_0}^{tot}}\dr\simeq\ga 0.43^{+ 1.07}_{-0.26}\dr~{\rm keV},
\eeq
 Using, ${\Gamma_{a_0}^{\eta\pi}}/
{\Gamma_{a_0}^{tot}}\simeq 0.093\pm 0.020$ and $M_{a_0}=1474$ MeV, one deduces:
\beq
\tilde g_{a_0\gamma\gamma}\simeq \ga 0.26\pm 0.14\dr~\rm{GeV}^{-1}~.
\eeq

 %%%%%%%%%%%%%%%%%%%%%%%%%%%%%%%%%%%%%%%
 \section{$a_\mu ^{lbl}\vert_S$  and comparison with some other evaluations}
\label{sec:a^lbl_values}
\setcounter{equation}{0}
 %%%%%%%%%%%%%%%%%%%%%%%%%%%%%%%%%%%%%%%
% \subsection*{\b Our estimate and comparison with some others}
 %%%%%%%%%%%%%%%%%%%%%%
 The scalar exchange contribution to the muon anomalous 
 magnetic moment is given by Eq.\,\rf{a_mu_integrals}.
% \beq
% a_\mu^{lbl}\simeq \sum_{S\equiv \sigma,\cdots} |\tilde g_{S\gamma\gamma}|^2{\cal I}_{S}
% \eeq
The integrals ${\cal I}_{p}$, ${\cal I}_{pq}$ , ${\cal I}_{q}$ have been evaluated numerically,
and their values are given in Table\,\ref{tab:amupp} versus the value of the scalar meson mass.
 %%%%%%%%%%%%%%%%%%%%%%%%%%%%%%%%%%%%%%%%%
Our results in Table \ref{tab:amupp}, which are shown for different values of $\kappa_S$, 
are expected to take into account all $S$-waves contributions (direct $\oplus$ rescattering) 
as we have used the {\it total} $\gamma\gamma$ widths for each meson. Before going over to the comparison
of our results with some of those already available in the literature,
let us make a few comments about the results shown in Table \ref{tab:amupp}:\\
\b As discussed at the end of Section \ref{sec:Gamma^S}, an analysis based only on the leading short-distance behavior of the vertex function $\Gamma_{\mu\nu}^S$ and on the VMD representation of the form factors does not properly account for the decay of pure isovector scalar states into two photons, whereas the analysis of Ref. \cite{MS} leads to the choice $\kappa_S=1$ in this case. Due to the possible mixing of the isoscalar mesons with gluonium states, the corresponding value of $\kappa_S$ cannot be fixed without further knowledge on the matrix elements in Eq. \rf{mathix_elements}, and will in general even be different for each scalar meson. In Table \ref{tab:amupp} we have considered two values of $\kappa_S$: $\kappa_S=0$, i.e. no contribution from the form factor ${\cal Q} (q_1,q_2)$, and $\kappa_S=1$, which follows from the analysis of Ref. \cite{MS}. \\
\b One can notice that the contributions from the $\sigma/f_0(500)$ to\,$a_\mu^{lbl}$
dominate over the other scalar contributions, independently of the value of $\kappa_S$. 
This dominance of the $\sigma$ contribution over the other scalar mesons can be understood, on the one hand, 
from the behaviour of the weight functions defined in Table\,\ref{tab:weight1} and shown in Figs\,\ref{fig:weight1}, 
\ref{fig:weight2} and \ref{fig:weight3} versus $Q_1^2$ and $Q_2^2$, which are more weighted, like
in the case of the pion exchange \cite{KN}, for the mesons of lower masses, and, on the other hand, 
by the fact that the $\gamma\gamma$ couplings of 
higher states are much smaller than the one of the $\sigma$.\\
\b The contributions of the higher-mass states $f_0(1370)$, $a_0(1450)$ and $f_0(1500)$
are not suppressed as compared to the lighter states $a_0(980)$ and $f_0(990)$ as could
naively be expected from a simple scaling argument of the masses. Another important parameter
here is the two-photon width. The coupling of the heavier scalars to a photon pair turns out
to be rather strong as compared to the light scalars.
\\
\b If we only consider the contribution from the 
Lorentz structure ${\cal P}_{\mu\nu}$ to the $\sigma\gamma\gamma$ form factor in Eq. (\ref{eq:invariant}),
like often done in the current literature, one obtains [case ${\cal Q}(0,0)=0$ in Table \ref{tab:amupp}]:
\bea
a_\mu^{lbl}\vert_\sigma&=&-\ga 5.35^{+3.27}_{-0.92}\dr\times 10^{-11}%,\nnb\\
%\sum_{S\equiv a_0,f_0,...}\hspace*{-0.35cm}a_\mu^{lbl}\vert_S&=&-(3.74^{+3.17}_{-0.70})\times 10^{-11}
,
\eea
where the $\sigma$ contribution is comparable in size and sign with the resuls obtained by
other authors \cite{Bijnens:1995xf,BARTOS} [the value given in Ref. \cite{RAF} is the same
as in Ref. \cite{Bijnens:1995xf}, but with the uncertainty scaled to 100\%],
and  with the one using $\pi\pi$ rescattering analysis \cite{Colangelo:2017qdm} quoted in Table \ref{tab:others},
with which some connection can be established from the methodological point of view.

%\noindent
This brings us to a more direct comparison with the results obtained by the authors
of Ref. \cite{PAUK} on the one hand, and of Refs. \cite{Colangelo:2017qdm,Colangelo:2017fiz}
on the other hand.\\
\b The authors of Ref. \cite{PAUK} consider the contribution
to HLbL coming from the scalar mesons $f_0(990)$, $a_0(980)$
and $f_0(1370)$ in the same NWA as considered here. They start from 
a different decomposition of the vertex function $\Gamma_{\mu\nu}^S$:
\be
\Gamma_{\mu\nu}^S = {\cal F}_{TT} T_{\mu\nu} + {\cal F}_{LL} L_{\mu\nu} 
,
\ee
which describes the production of a scalar meson,
for instance in $e^+ e^- \to e^+ e^- S \ (\to e^+ e^- \pi\pi)$, through
either two transverse or two longitudinal photons \cite{Poppe_86}.
The link with the decomposition in Eq. \rf{S_FF} is given by:
\bea
{\cal F}_{TT}(q_1 , q_2) \!\! &=& \!\! - (q_1 \cdot q_2) {\mathcal P} (q_1 , q_2) - q_1^2 q_2^2{\mathcal Q} (q_1 , q_2),
\nonumber\\
\\
{\cal F}_{LL}(q_1 , q_2) \!\! &=& \!\!  - (q_1 \cdot q_2) \left[
{\mathcal P} (q_1 , q_2) + (q_1 \cdot q_2){\mathcal Q} (q_1 , q_2)
\right]
.
\nonumber
\eea
In their analysis, they assume that the
contribution from the longitudinal part ${\cal F}_{LL}(q_1 , q_2)$
is suppressed [as compared to the one from ${\cal F}_{TT}(q_1 , q_2)$]
and thus they do not consider it. Moreover, they use, for the
transverse form factor, a monopole representation, which is reproduced
by the VMD representation used here when $B=0$, i.e. $\kappa_S=0$, a choice 
which then consistently also entails that ${\mathcal Q}^{\rm VMD} (q_1 , q_2)=0$ (see 
Eq. (\ref{eq:formfactor})). As shown by the results in Table 4, the contribution from the form factor ${\cal Q} (q_1,q_2)$ is in general substantial.

%In order to compare with the results of Ref. \cite{PAUK}, we need to
%move the mass $M_V$ from the value $M_V=775$ MeV used here
%to the range considered there for the monopole mass, namely 1 GeV$\le M_V\le$ 2 GeV.
%This then gives: 
%\be
%a_\mu ^{lbl}\vert_S^{\kappa_S=0}\times 10^{11}
%=
%\left\{
%\begin{tabular}{l}
%$-S~:~a_0(980)$\\
%$-S~:~f_0(990)$\\
%$-S~:~f_0(1370)$
%\end{tabular}
%\right.
%\ee
%
%%\be
%a_\mu ^{lbl}\vert_S^{\kappa_S=0}\times 10^{11}
%=
%\left\{
%\begin{tabular}{l}
%$-(0.43\pm 0.14)~:~a_0(980)$\\
%$-(0.42\pm 0.09)~:~f_0(990)$\\
%$-(0.24\pm 0.11)~:~f_0(1370)$
%\end{tabular}
%\right.
%\ee
%\\
%\bea
%a_\mu^{lbl}\vert_\sigma&=&-\ga 15.56\pm 5.3\dr\times 10^{-11},\nnb\\
%\sum_{S\equiv a_0,f_0,...}\hspace*{-0.35cm}a_\mu^{lbl}\vert_S&=&-(1.9\pm 0.2)\times 10^{-11}.
%-(5.75_{-1.16}^{+1.64})&\leq& a_\mu^{lbl}\vert_\sigma\times 10^{11}\leq + 14.19_{-2.99}^{+19.61},\nnb\\
%-(3.73^{+3.17}_{-0.70})&\leq&\hspace*{-0.35cm} \sum_{S\equiv a_0,f_0,...}\hspace*{-0.35cm}a_\mu^{lbl}\vert_S\times 10^{11}\leq -(0.30^{+1.80}_{-0.38}),\nnb\\
%\eea
\b In Refs. \cite{Colangelo:2017qdm,Colangelo:2017fiz}, the $\pi\pi$ rescattering effects
to HLbL are considered, with $\gamma^*\gamma^*\to\pi\pi$ helicity partial waves $h_{J;\lambda_1\lambda_2}$
[$\lambda_i$ denote the photon helicities] constructed dispersively, using $\pi\pi$ phase shifts 
derived from the inverse amplitude method. The $I=0$ part of this calculation, which gives:
\be
a_{\mu; J=0; I=0}^{\pi\pi;\pi-{\rm pole~LHC}} = -9 \cdot 10^{-11}
\lbl{amu_piLHC}
\ee 
with a precision of 10\%, can be interpreted as the contribution from the $\sigma/f_0(500)$ meson.
The mention ``$\pi-{\rm pole~LHC}$" means that the left-hand cut is provided by the Born term alone,
i.e. single-pion exchange in the $t$ channel.
Instead of $\Gamma_{\mu\nu}^S$, the starting point is the matrix element:
\be
\int d^4 x \, e^{i q_1 \cdot x} 
\langle \Omega \vert T \{ j_\mu (x) j_\nu (0) \} \vert \pi^a (p_1) \pi^b(p_2) \rangle
,
\ee
where either $a=b=0$, or $a=+$, $b=-$. These matrix element can be decomposed in
terms of four independent invariant functions $A_i$ in the following way
(see e.g. Ref. \cite{Colangelo15}):
\be
- A_1 P_{\mu\nu} (q_1 , q_2) - A_2 Q_{\mu\nu} (q_1 , q_2)\, + \!\! \sum_{i=3,4,5} A_i T^i_{\mu\nu} (q_1 , q_2)
,
\ee
where $p_1 + p_2 = q_1 + q_2$. The expressions of the remaining tensors $T^i_{\mu\nu} (q_1 , q_2)$
for $i=3,4,5$ are not needed here, and can be found in Ref. \cite{Colangelo15}. What matters is that, upon performing a
partial wave decomposition, only $A_1$ and $A_2$ receive contributions from the $S$ wave. In the NWA,
the vertex function $\Gamma_{\mu\nu}^S (q_1,q_2)$ arises as the residue of the pole as $s\equiv (q_1+q_2)^2 \to M_S^2$,
the correspondence being:
\be
h_{0,++} (s) \to - \frac{1}{4} {\cal F}_{TT} , 
\ h_{0,00} (s) \to - \frac{1}{4} \frac{\sqrt{q_1^2} \sqrt{q_2^2}}{(q_1 \cdot q_2)} {\cal F}_{LL}
.
\ee
In addition, the Born term in the $\pi^+\pi^-$ channel only contributes to $A_1$ and to $A_4$, which
in turn has no $J=0$ component, but not to $A_2$. There is therefore a 
relation between the Born term contributions to $h_{0,++}$ and to $h_{0,00}$,
which effectively amounts to the condition ${\cal Q}(q_1 , q_2)=0$, i.e. 
$\kappa_S = 0$. The result we obtain for this value (see Table \ref{tab:amupp})
is somewhat higher than the number quoted in Eq. \rf{amu_piLHC}, but this difference can possibly be understood by the absence of a more
complete description of the left-hand cut in the analysis of Refs. \cite{Colangelo:2017qdm,Colangelo:2017fiz}.
%%%%%%%%%%%%%%%%%%%%%%%%%%%%%%%%%%%%%%%
 {\scriptsize
\begin{table}[t]
\begin{center}
%\begin{table*}[hbt]
\setlength{\tabcolsep}{0.5pc}
%\newlength{\digitwidth} \settowidth{\digitwidth}{\rm 0}
%\catcode`?=\active \def?{\kern\digitwidth}

 \caption{\footnotesize    
Different estimates of the scalar meson contributions via LbL scattering at lowest order (LO).
We use $\Gamma_{\sigma}^{\gamma\gamma}=1.62(42)\,{\rm keV}$  in Eq.\,\ref{eq:sigma-gamma}.}
    {\footnotesize
\begin{tabular}{lcc}
\hline
\hline
\\
Scalar&$ a_\mu ^{lbl}\vert_S\times 10^{11}$&Refs \\
\\
\hline
\\
{\bf This work}\\
%\\
$\sigma(960\pm 96)$&$-\ga 3.14_{-0.72}^{+0.84}\dr\leq...\leq -\ga 0.31 ^{ +0.41}_{-0.82}\dr$&This work \\
\\
$\sum_{a_0,f_0,...}$&$-\ga 2.21^{+3.16}_{-0.65}\dr\leq...\leq -\ga 0.99^{+2.02}_{-0.40}\dr$&--\\
\\
Total sum &$-\ga 5.35^{+3.27}_{-0.92}\dr\leq ...\leq -\ga 1.3^{+2.06}_{-0.91}\dr$&--\\
\\
{\it Final result} &$-\ga 4.51\pm 4.12\dr$ & This work \\
%\it Total &$-(9.48^{+3.57}_{-1.35})\leq...\leq+(13.89_{-3.01}^{+19.69})$&This work\\
\\
{\bf Others} \\
$\sigma(620)$   &$-(6.8\pm 2.0)$&ENJL\,\cite{Bijnens:1995xf}\\
$\sigma(620)$&$-(6.8\pm 6.8)$ &ENJL\,\cite{RAF}\\
$\sigma(400-600)$&$-(36\sim 7)$&\cite{BARTOS}\\ %: $\Gamma_{\sigma}^{\gamma\gamma}=1.0\,{\rm keV}$\\
$\pi\pi$-rescattering &$-(7.8\pm 0.5)$&$\pi$ pole\,\cite{Colangelo:2017qdm} \\
\\
\hline\hline
\end{tabular}
%\end{tabular*}
}
\label{tab:others}
\end{center}
\end{table}
}
%\vspace*{-1cm}
 %%%%%%%%%%%%%%%%%%%%%%%%%%%%%%%%%%%
 {\scriptsize
\begin{table}[ht]
\begin{center}
%\begin{table*}[hbt]
\setlength{\tabcolsep}{2.1pc}
 \caption{\footnotesize    
Recent determinations of the LO hadron vacuum polarization (HVP) in units of $10^{-11}$
from the data compared with some other models and lattice results. The tentative theoretical average is more weighted by the most precise determinations in\,\cite{NOMURA18,EDUARDO17}. The weighted averaged error is informative. Instead, one may use the one from the precise determinations which is about twice the averaged error.}
\vspace*{0.25cm}
    {\footnotesize
\begin{tabular}{lll}
\hline
\hline
\\
Values&Refs \\
\\
\hline
\\
{\bf Data}\\
%!6949.1&$\pm$42.7&\cite{HAGIWARA11}\\
6880.7$\pm$41.4&\cite{JEGER15}\\
6931$\pm$34&\cite{DAVIER17} \\
6933$\pm$25&\cite{NOMURA18} \\

\it  6922.4$\it\pm18.1$&\hspace{-0.6cm} \it Data Average  \\
\\
{\bf Models $\oplus$ Lattice data}\\
6932$\pm$25& \cite{EDUARDO17}\\
6818$\pm$31& \cite{BENAYOUN16}\\
6344$\pm$354&  \cite{DOMINGUEZ17} \\
\\
%\it  6885.4&\it$\it\pm19.4$&\hspace{-0.85cm} \it Model Average \\
{\bf Lattice}\\
6740$\pm$277& \cite{ETM} \\
6670 $\pm$134.2&\cite{HPQCD}\\
7110 $\pm$188.6&\cite{BMW} \\
6540$\pm$388&\cite{DELLA} \\
7154$\pm$ 187&\cite{BLUM}\\
6830$\pm$180&\cite{GIUSTI} \\
%\it  6810.1&\it$\it\pm101.7$&\hspace{-0.9cm} \it Lattice Average \\
\\
\bf Tentative Theoretical Average\\
 6904.02$ \pm 13.06$&\\ %\hspace{-1.4cm}\it Tentative Theoretical Average \\
\\
\hline\hline
\end{tabular}
}
\label{tab:hvp}
\end{center}
\end{table}
}
%\vspace*{-1cm}
%\vspace*{-1cm}

%%%%%%%%%%%%%%%%%%%%%%%%%%%%%%%%%%%
 {\scriptsize
\begin{table}[H]
\begin{center}
%\begin{table*}[hbt]
\setlength{\tabcolsep}{0.75pc}
 \caption{\footnotesize    
Comparison  of the different determinations of the pseudoscalar meson contributions  in units of $10^{-11}$.
We have taken the mean of the asymmetric errors  in the average which is about 0.8 the one of the most precise error.  
}
\vspace*{0.25cm}
    {\footnotesize
\begin{tabular}{llll}
\hline
\hline
\\
Values&Approaches&Refs \\
\\
\hline
\\
83.0$\pm12.0$& Vector Meson Dominance&\cite{KN}\\
84.0$^{+8.7}_{-8.1}$&Vector Meson Dominance& \cite{NYF}\\
89.9$^{+9.7}_{-8.9}$&Lowest Meson Dominance $\oplus$ Vector& \cite{NYF}\\
84.7$^{+5.3}_{-1.8}$&Resonance Chiral Theory& \cite{CILLERO} \\
\\
$\it 85.0\pm 3.6$& &\it Average \\
\\
\hline\hline
\end{tabular}
}
\label{tab:lblp}
\end{center}
\end{table}
}
%\vspace*{-1cm}
%\vspace*{-1cm}

\indent

 %%%%%%%%%%%%%%%%%%%%%%%%%%%
 \section{Present Experimental and Theoretical Status}
\label{sec:status}
\setcounter{equation}{0}
 %%%%%%%%%%%%%%%%%%%%%%%%%%%
 We show in Table\,\ref{tab:hvp} the different estimates of $a_\mu^{hvp}$, where one may amazingly notice that
% from which we take 
 the mean of the two recent phenomenological determinations\,\cite{DAVIER17} 
 and \cite{NOMURA18} % from the data and retain the most precise error. This average amazingly 
 co\"\i ncides with the one obtained in\,\cite{EDUARDO17} whithin a theoretical model. 
 Using our new estimate of the scalar meson contributions to the Light-by-Light scattering to $a_\mu$,  we show in Table\,\ref{tab:status} the present experimental and theoretical  
 status on the determinations of $a_\mu$.

  %%%%%%%%%%%%%%%%%%%%%%%%%%%%%%%%%%%%%%%
 {\scriptsize
\begin{table}[H]
\begin{center}
%\begin{table*}[hbt]
\setlength{\tabcolsep}{.8pc}
%\newlength{\digitwidth} \settowidth{\digitwidth}{\rm 0}
%\catcode`?=\active \def?{\kern\digitwidth}
 \caption{\footnotesize    
Comparison  of the experimental measurement and theoretical determinations of $a_\mu$ within the Standard Model (SM) 
in units of $10^{-11}$. For HVP at LO, we take the tentative theoretical average obtained in Table\,\ref{tab:hvp}. For the pseudoscalars contributions to HLbL, we take the mean of the ones in Table\,\ref{tab:lblp}. For the scalars, we take the mean of the errors quoted in the final result of this work in Table\,\ref{tab:others}. The total errors of the sum in the present Table have been added quadratically.}
\vspace*{0.25cm}
    {\footnotesize
\begin{tabular}{lll}
\hline
\hline
\\
Determinations&Values &Refs \\
\\
\hline
\\
{\bf Experiment} &11~659~2091.0$\pm63.0$&\cite{EXP} \\
\\
{\bf Theory}\\
{\it QED at 5 loops}&11~658~4718.85$\pm 0.36$&\cite{KIN,KURZ}\\
{\it Electroweak at 2 loops}&$+(154.0\pm1.0)$ &\cite{KP,CM}\\
\\
{\it  HVP}%Hadronic Vacuum Polarisation}&{\it}
\\
LO &$+(6904.02\pm 13.06)$&Average \\
NLO &$-(99.34\pm 0.91)$& \cite{KURZ2,JEGER15}\\
N2LO &$+(12.26\pm 0.12)$&\cite{JEGER15}\\
\it Total HVP&$ +(6816.94\pm 13.09)$ \\
\\
%{\it Hadronic Light-by Light}&\it Scattering at LO \\
\it HLbL at LO \\
Pseudoscalars& $+\ga 85.0\pm 2.8\dr$&Average \\
%$\pi,~K$ loops &-19&16&\cite{PRADESb,RAF}\\
Scalars %LO $\oplus~(\pi,K)$ loops
&$-\ga 4.51\pm 4.12\dr$&This work \\
%&$\geq -8.80$ &2.27&This work\\
Axial-vector&$+( 7.5\pm 2.7)$&\cite{PAUK,JEGER15} \\
Tensor&$+(1.1\pm 0.1)$&\cite{PAUK}\\
\it Total HLbL&$+(88.0\pm 5.7)$\\
%\end{tabular}
\\
\boldmath $a_\mu^{\rm SM}$ &$11~659~1778.9 \pm 14.3 $& This work \\
\\
%\setlength{\tabcolsep}{.5pc}
%\begin{tabular}{llll}
%\hline\\
%{\it Theory: total sum} &$(11~659~1737.4\pm 44.2)$ & This work \\
\boldmath ${a_\mu^{\rm exp}-a_\mu^{\rm SM}}$ &$+(312.1\pm 64.6)$& This work \\
\\
\hline\hline
\end{tabular}
%\end{tabular*}
}
\label{tab:status}
\end{center}
\end{table}
}

 %%%%%%%%%%%%%%%%%%%%%%%
 \section{Conclusions}
\label{sec:concl}
\setcounter{equation}{0}
 %%%%%%%%%%%%%%%%%%%%%%%
 We have systematically studied the light scalar meson contributions to the anomalous magnetic moment of the muon $a_\mu$ from hadronic light-by-light scattering (HLbL). Our analysis also includes the somewhat heavier states, which however have couplings to two photons at least as strong as those of the $a_0(980)$ and the $f_0(990)$. Our results are summarized in Table \ref{tab:amupp} and compared with some other determinations in Table \ref{tab:others}. We conclude that the HLbL contribution from the scalars is dominated by the $\sigma/f_0$ one, which one may understand from the $Q^2$-behaviour of the weight functions entering into the analysis, and which are plotted in Figs.\,\ref{fig:weight1} to \ref{fig:weight3}. Moreover, the uncertainties on the parametrisation of the form factors induce large errors in the results, which might be improved from a better control of these observables. In particular, our analysis draws the attention to the potentially important contribution from the second structure ${\cal Q}_{\mu\nu}$  in the decomposition of the vertex function in Eq.\,\rf{S_FF}, which could even lead to a change of sign in $a_\mu^{lbl}\vert_\sigma$. For the isovector states, an estimate of its size could be obtained from the analysis of Ref. \cite {MS}. For the isoscalar states, mixing with glueball states and/or with ${\bar s}s$ states can lead to important contributions from the whole set of  matrix elements in Eq. \rf{mathix_elements}. Knowledge of these matrix elements can possibly be obtained, for instance, either from phenomenology or from QCD spectral sum rules. We leave this matter for a future research. For a conservative result, we consider as a (provisional) {\it final result} the range of values spanned by the two possible values from $0$ to 1 of ${\cal Q}(0,0)/(M_S^2\tilde g_{S\gamma\gamma})$ obtained in Table\,\ref{tab:amupp}, which we compare in Table\,\ref{tab:others} with some other determinations. Finally, we present in Table\,\ref{tab:status} a new comparison of the data with theoretical predictions including our new results. The theoretical errors from HLbL are dominated by the ones due to the scalar meson contributions. Moreover, some other scalar meson contributions to $a_\mu$ from radiative decays of vector mesons and virtual exchange  have also been  
considered in\,\cite{SNSCALAR}. We plan to improve these results in a future work.

\vspace*{-0.5cm}

%\vfill\eject
 %%%%%%%%%%%%%%%%%%%%%%%
 \section*{Acknowledgements}
 %%%%%%%%%%%%%%%%%%%%%%%
S. Narison wishes to thank Wolfgang Ochs for some useful correspondences, U. Gastaldi for several discussions on the scalar mesons, and the ICTP-Trieste for a partial financial support. M. Knecht wishes to thank P. Sanchez-Puertas for useful remarks on the manuscript. The work of M.K. and S. N. has been carried out thanks to the support of the OCEVU Labex (ANR-11-LABX-0060) and the A*MIDEX project (ANR-11-IDEX-0001-02) funded by the “Investissements d'Avenir” French government program managed by the ANR. 

% \vspace{0.1cm}
 
 %%%%%%%%%%%%%%%%%%%%%%%%
% \vfill\eject

\end{document}